\documentclass[12pt,showpacs,prd]{revtex4}
\usepackage{epsfig,amssymb,amsfonts}

\makeatletter
\renewcommand{\@makefntext}[1]{\parindent=1em\noindent\hbox to 1.8em{\hss$^{\@thefnmark}$}#1}
\renewcommand{\@footnotemark}{\hbox{\mathsurround=0pt$^{\@thefnmark}$}}
\newcommand{\ftnote}[2]{\footnotemark[#1]\footnotetext[#1]{#2}}
\makeatother

\begin{document}
\title{Mesonic states and vacuum replicas in potential quark models for
QCD}\author{A. V. Nefediev}
\affiliation{Institute of Theoretical and Experimental Physics, 117218,\\
B.Cheremushkinskaya 25, Moscow, Russia}
\author{J. E. F. T. Ribeiro}
\affiliation{ Centro de F\'\i sica das Interac\c c\~oes Fundamentais
(CFIF),Departamento de F\'\i sica, Instituto Superior T\'ecnico, Av.
Rovisco Pais, P-1049-001 Lisboa, Portugal}
\newcommand{\be}{\begin{equation}}
\newcommand{\bea}{\begin{eqnarray}}
\newcommand{\ee}{\end{equation}}
\newcommand{\eea}{\end{eqnarray}}
\newcommand{\ds}{\displaystyle}
\newcommand{\low}[1]{\raisebox{-1mm}{$#1$}}
\newcommand{\loww}[1]{\raisebox{-1.5mm}{$#1$}}
\newcommand{\lmn}{\mathop{\sim}\limits_{n\gg 1}}
\newcommand{\vpint}{\int\makebox[0mm][r]{\bf --\hspace*{0.13cm}}}
\newcommand{\too}{\mathop{\to}\limits_{N_C\to\infty}}
\newcommand{\vp}{\varphi}

\begin{abstract}
In this paper, we study generic non-local NJL models for
four-dimensional QCD in the limit of a large number of colours. We
diagonalise, through a Bogoliubov-type transformation, the
Hamiltonian of the model completely in terms of the compound
operators creating/annihilating mesons --- bound states of dressed
quarks and antiquarks --- and demonstrate a one-to-one
correspondence between the Bogoliubov diagonalisation condition
and the Bethe--Salpeter equation for bound states. We conclude
that, in the leading order in $N_C$, the nontrivial contents of
the theory is entirely encoded in the chiral angle --- the
solution to the mass-gap equation for dressed quarks.  As a
consequence, we extend the statement of existence of excited
solutions to the mass-gap equation, called vacuum replicas, beyond
the BCS level and prescribe a new index to the mesonic operators
--- the label of the vacuum state in which the
corresponding meson is created.
\end{abstract}
\maketitle

\section{Introduction}

For nearly half a century Nambu--Jona-Lasinio (NJL)\cite{NJL}
class of models \cite{Orsay,Orsay2,Lisbon} have been quite useful
tools for the on going study and understanding of low-energy
phenomena in QCD. Quark models with non-local quark
current-current microscopic interactions, parameterised by kernels
of particular forms, constitute prominent steps towards building
an effective low-energy theory for the strong interactions.
Besides regularising the ultraviolet divergences of the theory,
these nonlocal quark kernels bring in the necessary interaction
scale needed to make contact with the hadronic phenomenology.
Models with quark instantaneous interaction modeled after the
harmonic oscillator kernels \cite{Orsay,Orsay2,Lisbon} or kernels
obtained in the Coulomb-gauge QCD \cite{coulg}, as well as a
variety of other approaches based on the linear confinement
\cite{linear}, constitute as many examples of closely related
approaches exploiting the idea of chirally nonsymmetric solutions
of a nonlinear gap equation. In this paper, we continue our
studies of quark models \cite{Lisbon} with instantaneous
inter-quark interaction. An important feature of this class of
models is the essential covariance of results and conclusions
across a variety of possible forms for the confining kernel. For
example, the pion-pion low-energy scattering problem can be solved
in the general form, in terms of the pion mass and decay constant,
regardless of the explicit form of the quark kernel, provided it
is confining \cite{pipi}. Another example of such universality is
given by the existence of multiple chirally non-invariant
solutions to the mass-gap equation, predicted in
\cite{replica1,replica2}, found both analytically and numerically
for all allowed power-like confining potentials in
\cite{replica4}, and independently confirmed by calculations done
in a different approach in \cite{replica3}. The same conclusion of
universality holds for any low-energy phenomena considered in the
framework of such class of models and, therefore, we dare to
expect to be able to reproduce the low-energy properties of QCD.
An important progress was achieved in \cite{2d} where a
two-dimensional model for QCD \cite{tHooft} in the axial gauge was
considered and a second, Bogoliubov-type, transformation was
performed in the mesonic sector to fully diagonalise, up to
corrections of order $O(1/\sqrt{N_C})$, the Hamiltonian of the
theory in terms of the mesonic creation/annihilation operators.
The Hamiltonian approach to two-dimensional QCD in the light-cone
gauge was also developed in \cite{japan}. In the present paper, we
extend this result to four-dimensional QCD and establish a
one-to-one correspondence between the Bogoliubov condition (which
ensures that the Hamiltonian is diagonalised in the mesonic
sector) and the Bethe--Salpeter equation for bound quark-antiquark
states. We build the corresponding amplitudes explicitly and study
in detail the cases of the chiral pion and of the $\rho$-meson. We
conclude that solving the Bethe--Salpeter equation in the ladder
(rainbow) approximation is equivalent, in the leading order in
$N_C$, to go beyond BCS level and to diagonalise the quartic term
in the Hamiltonian of the theory. This feature is general and
holds for any confining quark kernel. We use the oscillator-type
quark kernel just as an example to illustrate the method. We also
conclude that all the information needed for the second and final
mesonic Bogoliubov transformation, is already contained in the
solutions of the mass-gap equation appearing at the BCS level.
Therefore, in the leading order in $N_C$, the chiral angle --- the
solution to the mass-gap equation
--- remains the only nontrivial characteristic of the theory and
it defines the latter completely. As a result, we discuss the
multiple chirally noninvariant solutions of mass-gap equations
--- the vacuum replicas \cite{replica1,replica2,replica4} --- and
conclude that, once such solutions exist, this information gets
carried beyond BCS and appears as an extra spectroscopic index for
mesonic states on top of the usual set $\{ {\vec P},n,J^{PC}\}$---
namely, each meson carries the following quantum numbers: the
total momentum ${\vec P}$, the radial excitation number $n$, the
$J^{PC}$ set, and, finally, the label of the vacuum state
in which it is created. We discuss possible consequences of this
generalisation.

The paper is organised as follows. The second section is first
devoted to the diagonalisation of the Hamiltonian in the
quark sector (subsection A) and then to its subsequent
diagonalisation in the mesonic sector (subsection B). We end this
chapter with the discussion of the equivalence between the mesonic
Bogoliubov-Valatin transformation and the Bethe--Salpeter
equation. Vacuum replicas are studied in the third chapter. We
give the necessary details of the theory of replicas, in the first
subsection, and comment on the tachyon problem, in the second
subsection. Finally, in the third subsection, we generalise the
theory of replicas beyond BCS and discuss the hadronic spectrum in
the replica vacua. The fourth and last chapter of the paper is
devoted to our conclusions and outlook.

\section{Diagonalisation of the quark Hamiltonian}
\subsection{Diagonalisation in the single-quark sector and the mass-gap
equation}

In this section, we give the necessary details of the quark model
\cite{Orsay,Orsay2,Lisbon} to be used in this paper and describe
the Hamiltonian diagonalisation in the single-quark sector (BCS
level). Naively one should expect this to be the end of the story.
However, as we shall see, the requirement of confinement, that is,
no free asymptotic quarks may be allowed, will completely change
this picture and we shall end up with a diagonal Hamiltonian
written in terms of asymptotic mesons rather than asymptotic
quarks. How this comes about will be the subject of this paper.

The Hamiltonian of the model,
\be
\hat{H}=\int d^3 x\bar{\psi}(\vec{x},t)\left(-i\vec{\gamma}\cdot
\vec{\bigtriangledown}+m\right)\psi(\vec{x},t)+ \frac12\int d^3
xd^3y\;J^a_\mu(\vec{x},t)K^{ab}_{\mu\nu}(\vec{x}-\vec{y})J^b_\nu(\vec{y},t),
\label{H}
\ee
contains the interaction of the quark currents
$J_{\mu}^a(\vec{x},t)=\bar{\psi}(\vec{x},t)\gamma_\mu\frac{\lambda^a}{2}
\psi(\vec{x},t)$ parameterised through the instantaneous quark
kernel,
\be
K^{ab}_{\mu\nu}(\vec{x}-\vec{y})=g_{\mu 0}g_{\nu 0}\delta^{ab}
V_0(|\vec{x}-\vec{y}|).
\label{KK}
\ee
In Eq.~(\ref{KK}) we have used the simplest form of the kernel
compatible with the requirement of confinement. Spatial components
can be also included in the kernel (\ref{KK}) \cite{Lisbon}. They
affect, for example, the form of the mass-gap equation for the
chiral angle and spin-dependent terms in the effective quark-quark
interaction after a Foldy--Wounthuysen transformation (see, for
example, papers \cite{hl} where a heavy-light quarkonium was
considered in the modified Fock-Schwinger gauge and an effective
inter-quark interaction was derived).

But it is a fact, already mentioned in the introduction, that
despite all the aforementioned detail variations of possible quark
kernels, they all share with the kernel (\ref{KK}) the {\emph
same} low energy properties (Goldstone pion, $\pi -\pi$ Weinberg
scattering lengths, Gell-Mann--Oakes--Renner formula, and so on) so that it
is sufficient to stick to the form (\ref{KK}) as the mainstay of
this work and comment on the generalisations when it is the case.
Finally it is sufficient for phenomenological applications that
$V_0(|\vec{x}|)$ should interpolate between long range confinement
and short-range Coulomb behaviour. An example of such a
phenomenological kernel which is able to address both, heavy-quark
and light-quark limits of the theory, was considered in detail in
\cite{replica1}. A pure confining kernel of the generalised
power-like form,
\be
V_0(|\vec{x}|)=K_0^{\alpha+1}|\vec{x}|^{\alpha},
\label{potential}
\ee
with $K_0$ being the mass parameter, was studied for the first
time, in the series of papers \cite{Orsay} and an analysis was performed
concerning the general properties of the mass-gap equations for
such potentials. This analysis was extended in a recent paper
\cite{replica4} and the range of possible powers $\alpha$ was
identified to be $0\leqslant\alpha\leqslant 2$. In what follows,
we do not need to specify any particular form for $V_0(|\vec{x}|)$
other than requiring it to be confining. When needed, we shall use
the oscillator-type quark kernel corresponding to the case of
$\alpha=2$ in Eq.~(\ref{potential}), to exemplify some general
results. Convenience of such a choice follows from the fact that
the Fourier transform of the potential $\vec{x}^2$ is given by the
Laplacian of the three-dimensional delta function, so that all
integral equations can be transformed into differential equations
which are much easier for analytical and numerical studies
\cite{Orsay,Orsay2,Lisbon}.

Now we proceed in the standard way by defining dressed quarks with
the help of a Bogoliubov-Valatin transformation having the chiral
angle $\vp(p)\equiv\vp_p$ \cite{Orsay,Lisbon} as a parameter:
\be
\psi_{\alpha}(\vec{x},t)=\sum_{s=\uparrow,\downarrow}\int\frac{d^3p}{(2\pi)
^3}e^{i\vec{p}\vec{x}} [\hat{b}_{\alpha
s}(\vec{p},t)u_s(\vec{p})+\hat{d}_{\alpha s}^\dagger(-\vec{p},t)
v_s(-\vec{p})],
\label{psi}
\ee
\be \left\{
\begin{array}{rcl}
u(\vec{p})&=&\frac{1}{\sqrt{2}}\left[\sqrt{1+\sin\vp_p}+
\sqrt{1-\sin\vp_p}\;(\vec{\alpha}\hat{\vec{p}})\right]u(0),\\
v(-\vec{p})&=&\frac{1}{\sqrt{2}}\left[\sqrt{1+\sin\vp_p}-
\sqrt{1-\sin\vp_p}\;(\vec{\alpha}\hat{\vec{p}})\right]v(0),
\end{array}
\right.
\label{uandv}
\ee
\be
\hat{b}_{s}(\vec{p},t)=e^{iE_pt}\hat{b}_{s}(\vec{p},0),\quad
\hat{d}_{s}(-\vec{p},t)=e^{iE_pt}\hat{d}_{s}(-\vec{p},0),
\label{bandd}
\ee
where $E_p$ stands for the dispersive law of the dressed quarks;
$\alpha$ being the colour index, $\alpha=\overline{1,N_C}$. It is
convenient to define the chiral angle varying in the range
$-\frac{\pi}{2}<\vp_p\leqslant\frac{\pi}{2}$ with the boundary
conditions $\vp(0)=\frac{\pi}{2}$, $\vp(p\to\infty)\to 0$. Notice
that this definition is not unique and in some cases one is forced
to choose $\vp(0)=-\frac{\pi}{2}$, as we shall see.

The normal ordered Hamiltonian (\ref{H}) can be split into three parts, 
\be 
\hat{H}=E_{\rm vac}+:\hat{H}_2:+:\hat{H}_4:, 
\label{H3}
\ee 
and the usual procedure is to demand the quadratic part
$:\hat{H}_2:$ to be diagonal or, equivalently, that the vacuum
energy $E_{\rm vac}$ should become a minimum. Then, the
corresponding mass-gap equation ensures the anomalous Bogoliubov
terms $\hat{b}^\dagger \hat{d}^\dagger-\hat{d}\hat{b}$ to be absent in $:\hat{H}_2:$,
\be
A_p\cos\vp_p-B_p\sin\vp_p=0,
\label{mge}
\ee
where
\be
A_p=m+\frac12C_F\int\frac{d^3k}{(2\pi)^3}V_0(\vec{p}-\vec{k})\sin\vp_k,
\label{A} 
\ee 
\be 
B_p=p+\frac12C_F\int \frac{d^3k}{(2\pi)^3}\;
(\hat{\vec{p}}\hat{\vec{k}})V_0(\vec{p}-\vec{k})\cos\vp_k,
\label{B}
\ee
$C_F=\frac12(N_C-1/N_C)$ being the $SU(N_C)_C$ Casimir operator in
the fundamental representation. The large-$N_C$ limit implies that
the product $C_FV_0$ remains finite as $N_C\to\infty$, so that,
for the power-like form (\ref{potential}), an appropriate
rescaling of the potential strength $K_0$ is understood. In the
remainder of this paper we shall absorb the coefficient $C_F$ into
the definition of the potential, $C_FV_0(|\vec{x}|)\equiv
V(|\vec{x}|)$ by a trivial redefinition of $K_0$.

As soon as the mass-gap equation is solved and a nontrivial chiral
angle is found, the Hamiltonian (\ref{H3}) takes a diagonal form,
\be
\hat{H}=E_{\rm vac}+\sum_\alpha\sum_{s=\uparrow,\downarrow}\int
\frac{d^3 p}{(2\pi)^3} E_p[\hat{b}^\dagger_{\alpha s}(\vec{p}) \hat{b}_{\alpha
s}(\vec{p})+\hat{d}^\dagger_{\alpha s}(-\vec{p}) \hat{d}_{\alpha
s}(-\vec{p})],
\label{H2diag}
\ee
and the contribution of the $:\hat{H}_4:$ part is suppressed as
$1/\sqrt{N_C}$. The dressed quark dispersive law becomes
\be
E_p=A_p\sin\vp_p+B_p\cos\vp_p,
\label{Ep}
\ee
and this completes the diagonalisation of the Hamiltonian in the
quark sector to order $1/\sqrt{N_C}$ (the BCS level).

For further convenience we write out both the mass-gap equation
and the dressed quark dispersive law $E_p$ for the harmonic
oscillator potential, $V(|\vec{x}|)=K_0^3|\vec{x}|^2$
\cite{Orsay,Orsay2,Lisbon},
\be
p^3\sin\vp_p=\frac12K_0^3\left[p^2\vp''_p+2p\vp_p'+\sin2\vp_p\right]+
mp^2\cos\vp_p,
\label{diffmge}
\ee
\be
E_p=m\sin\vp_p+p\cos\vp_p-K_0^3\left[\frac{{\vp'_p}^2}{2}
+\frac{\cos^2\vp_p}{p^2}\right].
\label{Epharm}
\ee
In the remainder of this subsection, if not stated otherwise, we
consider the chiral limit $m=0$.

We end this introductory section by stating two general properties
held by mass-gap equations, Eq.~(\ref{mge}). First of all,
notice that using the explicit form of the dressed quark
dispersive law $E_p$ (\ref{Epharm}), the mass-gap equation
(\ref{diffmge}) can be rewritten in the form of a
Schr{\"o}dinger-like equation with the zero eigenvalue,
\be
[-K_0^3\Delta_p+2E_p]\psi=0,
\label{mgho}
\ee
where $\psi\equiv\sin\vp_p$. The generic mass-gap equation (\ref{mge})
for an arbitrary power-like confining potential (\ref{potential})
admits the form similar to (\ref{mgho}) \cite{replica4}.

The second general property concerns the asymptotic behaviour of
the solutions of the mass gap equation (the chiral angle) at large
momenta. Taking the general form of the mass-gap equation for the
power-like potential (\ref{potential}) \cite{replica4},
\be
p^3\sin\vp_p=K_0^{\alpha+1}\Gamma(\alpha+1)\sin\frac{\pi\alpha}{2}
\int_{-\infty}^{\infty}
\frac{dk}{2\pi}\left\{\frac{pk\sin[\vp_k-\vp_p]}{|p-k|^{\alpha+1}}+
\frac{\cos\vp_k\sin\vp_p}{(\alpha-1)|p-k|^{\alpha-1}}\right\},
\label{mg2}
\ee
and performing an expansion for $p\to\infty$, we get
\ftnote{1}{The case of the harmonic oscillator potential,
$\alpha=2$, has to be considered separately. As immediately
follows from Eq.~(\ref{mgho}) and the free limit of the quark
dispersive law, $E_p\mathop{\approx}\limits_{p\to\infty}p$, the
asymptotic behaviour of the chiral angle in this case is given by
the Airy function. Still the qualitative conclusion made below
holds true.}
\be
\left.\vp_p\right|_{m=0}\mathop{\approx}\limits_{p\to\infty}-
\frac{\pi}{N_C}\Gamma(\alpha+2)K_0^{\alpha+1}\sin\frac{\pi\alpha}{2}
\frac{\langle\bar{q}q\rangle}{p^{\alpha+4}}.
\label{asym}
\ee
From the behaviour (\ref{asym}) we see that, in the chiral limit,
the sign of the solution $\vp_p$ at large momenta is opposite to
the sign of the chiral condensate. Beyond the chiral limit the
expression (\ref{asym}) defines how the chiral angle approaches
the free large-momentum asymptote $\arctan\frac{m}{p}$. From the
Gell-Mann-Oakes-Renner formula \cite{GMOR} one can relate the
chiral condensate with the pion mass $M_\pi$,
\be
f_\pi^2M_\pi^2=-2m\langle\bar{q}q\rangle,
\label{GMOR}
\ee
so that, according to Eq.~(\ref{asym}), $M_\pi^2$ becomes positive
if the chiral angle approaches its large-momentum asymptote from
above, and negative if it approaches from below. In the latter
case the pion becomes a tachyon and this situation requires a
special treatment which will be discussed in detail in chapters II
and III.

\subsection{Diagonalisation in the mesonic sector and the bound-state
equation}

\subsubsection{Quark Hamiltonian in the space of quark-antiquark pairs}

In this subsection, we go beyond BCS level and include the quartic
part $:\hat{H}_4:$ of the Hamiltonian (\ref{H3}) into
consideration.

After diagonalisation at BCS level, we could have used the formal
invariance, under arbitrary Bogoliubov-Valatin transformations
(BV), of the quark field $\psi_{\alpha}(\vec{x},t)$ --- here
understood as an inner product between a Hilbert space, spanned
by the spinors $u_s(\vec{p})$ and $v_s(-\vec{p})$, and a Fock
space, spanned by the quark annihilation and creation operators.
This invariance requires that any given BV rotation in the Fock
space, must engender a corresponding counter-rotation in the
associated spinorial Hilbert space so as to maintain the inner
product $\psi_{\alpha}(\vec{x},t)$ invariant. Thence, the new
BV-rotated spinors will carry the information on the chiral angle
and, therefore, so does the quark propagators. Equipped with these
new Feynman rules, we can then proceed to evaluate a wealth of
physical processes among which the Bethe-Salpeter equations for
bound states play a central role. Formally, the Hamiltonian of
Eq.~(\ref{H}), when written in terms of the quark fields
$\psi_{\alpha}(\vec{x},t)$, is invariant under such BV
transformations and so it will contain --- through its $:\hat{H}_4:$
term --- Bogoliubov anomalous terms of various types.

{\em It will turn out that the set of all these Bethe-Salpeter
equations, each one of them associated with a corresponding
mesonic bound state, with their right mixture of positive with
negative-energy amplitudes, are exactly what is required in order
to get rid all the Bogoliubov anomalous terms with just four quark
creation(annihilation) operators. The remaining Bogoliubov terms
(responsible for generic hadronic decays like $A\to B+C$) will be
suppressed as $1/\sqrt{N_C}$}. This scenario is quite plausible
once we realise that a group of four-quark creation(annihilation)
operators --- two quarks and two antiquarks --- corresponds to a
bilinear creation (annihilation) of a group of two mesons and, as
we know, these bilinear anomalous terms can be made to disappear
by the use of an appropriated bosonic Bogoliubov-Valatin canonical
transformation.

For the implementation of the diagonalisation program we shall
follow the method of refs.~\cite{2d} generalising it to the
four-dimensional case. The basic idea of the method is the
following. At BCS level, we consider quarks as though they were
free, with confinement to be implemented at later stages, when
building the Bethe--Salpeter
equation for bound states. Instead we now start with the {\em ab-initio}
confinement requirement that only quark-antiquark pairs are
allowed to be created and annihilated, not single quarks, and that
the full Hamiltonian (\ref{H3}) must take a diagonal form in terms
of the operators creating/annihilating whole quark-antiquark
mesons. Therefore, every time a quark (antiquark) is created or
annihilated, an accompanying antiquark (quark) must be
created/annihilated as well. To write this {\em ab-initio} condition,
let us introduce a set of four colourless operators: two operators
which count the number of quarks and the number of antiquarks, 
\be
\begin{array}{c}
\hat{B}_{ss'}(\vec{p},\vec{p}')=\frac{\ds
1}{\ds\sqrt{N_C}}\sum_\alpha \hat{b}_{\alpha
s}^\dagger(\vec{p})\hat{b}_{\alpha s'}(\vec{p}'),\quad
\hat{D}_{ss'}(\vec{p},\vec{p}')=\frac{\ds
1}{\ds\sqrt{N_C}}\sum_\alpha \hat{d}_{\alpha
s}^\dagger(-\vec{p})\hat{d}_{\alpha s'}(-\vec{p}'),
\end{array}
\label{operatorsBD}
\ee
and two operators which create and annihilate a quark-antiquark pair,
\be
\begin{array}{c}
\hat{M}^\dagger_{ss'}(\vec{p},\vec{p}')=\frac{\ds
1}{\ds\sqrt{N_C}}\sum_\alpha \hat{b}^\dagger_{\alpha
s'}(\vec{p}')\hat{d}^\dagger_{\alpha s}(-\vec{p}),\quad
\hat{M}_{ss'}(\vec{p},\vec{p}')=\frac{\ds
1}{\ds\sqrt{N_C}}\sum_\alpha \hat{d}_{\alpha s}(-\vec{p})\hat{b}_{\alpha
s'}(\vec{p}').
\end{array}
\label{operatorsMM}
\ee
The coefficients in the definitions
(\ref{operatorsBD}) and (\ref{operatorsMM}) are fixed in such a
way that, in the large-$N_C$ limit, all four operators obey the
standard bosonic algebra, with the only nonzero commutator being
\be
[\hat{M}_{ss'}(\vec{p},\vec{p}')\;\hat{M}_{\sigma\sigma'}^\dagger(\vec{q},\vec{q}')]=
(2\pi)^3\delta^{(3)}(\vec{p}-\vec{q})
(2\pi)^3\delta^{(3)}(\vec{p}'-\vec{q}')\delta_{s\sigma}\delta_{s'\sigma'}.
\label{MMcom}
\ee

The pair (\ref{operatorsBD}) is sufficient to express the leading
part of the Hamiltonian (\ref{H2diag}),
\be
\hat{H}=E_{\rm
vac}+\sqrt{N_C}\sum_{s=\uparrow,\downarrow}\int\frac{d^3
p}{(2\pi)^3}E_p[\hat{B}_{ss}(\vec{p},\vec{p})+\hat{D}_{ss}(\vec{p},\vec{p})],
\label{H2diag2}
\ee
whereas the suppressed part $:\hat{H}_4:$ requires all four
operators.

According to our {\em ab-initio} confinement requirement, the operators
$\hat{B}$ and $\hat{D}$ cannot be independent and have to be
expressed through the operators $\hat{M}^\dagger$ and $\hat{M}$.
One easily builds such relations to be \cite{2d} 
\be 
\left\{
\begin{array}{c}
\hat{B}_{ss'}(\vec{p},\vec{p}')=\frac{\ds
1}{\ds\sqrt{N_C}}\ds\sum_{s''} \int\frac{\ds d^3p''}{\ds
(2\pi)^3}\hat{M}_{s''s}^\dagger(\vec{p}'',\vec{p})
\hat{M}_{s''s'}(\vec{p}'',\vec{p}')\\
\hat{D}_{ss'}(\vec{p},\vec{p}')=\frac{\ds
1}{\ds\sqrt{N_C}}\ds\sum_{s''} \int\frac{\ds d^3p''}{\ds
(2\pi)^3}\hat{M}_{ss''}^\dagger(\vec{p},\vec{p}'')
\hat{M}_{s's''}(\vec{p}',\vec{p}'').
\end{array}
\right.
\label{anzatz}
\ee

The anzatz (\ref{anzatz}) is the most crucial point of the method.
Formally, one can also view it as another solution to the set of
equations given by the commutators of the operators
(\ref{operatorsBD}) and (\ref{operatorsMM}). Notice that the
relations (\ref{anzatz}) correspond to the \lq\lq minimal
substitution", that is, each quark is accompanied by only one
antiquark, and {\em vice versa}, whereas, in the general case, a
whole quark-antiquark cloud should be also created together with
the accompanying particle. Since the creation of each next
quark-antiquark pair in the cloud is suppressed by an extra power
of $1/N_C$, then Eq.~(\ref{anzatz}) represents just the leading
terms of the full expressions.

After substitution of the anzatz (\ref{anzatz}) into the
Hamiltonian (\ref{H3}) we have no longer a general suppression of
the quartic part of the Hamiltonian as compared to the quadratic
part. Only a number of terms in $:\hat{H}_4:$ remain suppressed
and will be omitted henceforth.  The resulting Hamiltonian, in the
space of colourless quark-antiquark pairs, takes the form:
\be
\hat{H}=E_{\rm vac}'+\int \frac{d^3P}{(2\pi)^3}
\hat{\cal H}(\vec{P}),
\label{HH1}
\ee
where we have separated out the quark-antiquark cloud
centre-of-mass motion.

For the sake of simplicity, let us consider the Hamiltonian
density $\cal H$ at rest ($\vec{P}=0$),
$$
\hat{\cal H}\equiv\hat{\cal H}(\vec{P}=0)=\sum_{s_1s_2}\int\frac{d^3
p}{(2\pi)^3}
2E_p\hat{M}_{s_1s_2}^\dagger(\vec{p},\vec{p})\hat{M}_{s_2s_1}(\vec{p},\vec{p})
+\frac12\sum_{s_1s_2s_3s_4}\int\frac{d^3p}{(2\pi)^3}\frac{d^3q}{(2\pi)^3}
V(\vec{p}-\vec{q})
$$
\be
\times \left\{[v^{++}(\vec{p},\vec{q})]_{s_1s_3s_4s_2}
\hat{M}^\dagger_{s_2s_1}(\vec{p},\vec{p})\hat{M}_{s_4s_3}(\vec{q},\vec{q})\right.
+[v^{+-}(\vec{p},\vec{q})]_{s_1s_3s_4s_2}
\hat{M}^\dagger_{s_2s_1}(\vec{q},\vec{q})\hat{M}^\dagger_{s_3s_4}(\vec{p},\vec{p})
\label{HH2}
\ee
$$
\left.+[v^{-+}(\vec{p},\vec{q})]_{s_1s_3s_4s_2}
\hat{M}_{s_1s_2}(\vec{p},\vec{p})\hat{M}_{s_4s_3}(\vec{q},\vec{q})
+[v^{--}(\vec{p},\vec{q})]_{s_1s_3s_4s_2}
\hat{M}_{s_3s_4}(\vec{p},\vec{p})\hat{M}_{s_1s_2}^\dagger(\vec{q},\vec{q})\right\},
$$
with amplitudes $v$ given by
\be
\begin{array}{c}
[v^{++}(\vec{p},\vec{q})]_{s_1s_3s_4s_2}=
[u^\dagger_{s_1}(\vec{p})u_{s_3}(\vec{q})]
[v^\dagger_{s_4}(-\vec{q})v_{s_2}(-\vec{p})],\\[0cm]
[v^{+-}(\vec{p},\vec{q})]_{s_1s_3s_4s_2}=
[u^\dagger_{s_1}(\vec{p})v_{s_3}(-\vec{q})]
[u^\dagger_{s_4}(\vec{q})v_{s_2}(-\vec{p})],\\[0cm]
[v^{-+}(\vec{p},\vec{q})]_{s_1s_3s_4s_2}=
[v^\dagger_{s_1}(-\vec{p})u_{s_3}(\vec{q})]
[v^\dagger_{s_4}(-\vec{q})u_{s_2}(\vec{p})],\\[0cm]
[v^{--}(\vec{p},\vec{q})]_{s_1s_3s_4s_2}=
[v^\dagger_{s_1}(-\vec{q})v_{s_3}(-\vec{p})]
[u^\dagger_{s_4}(\vec{p})u_{s_2}(\vec{q})].
\end{array}
\label{ampls}
\ee

Although there are only two independent amplitudes in
(\ref{ampls}), for example, $v^{++}$ and $v^{+-}$, with the two
remaining amplitudes, $v^{--}$ and $v^{-+}$, easily related to the
first two by Hermitian conjugation and renaming of spin indices,
we prefer to keep all four amplitudes so as to give a more
symmetric form to both bound-state equations and the effective
diagrammatic rules to be derived shortly.

Notice that in general, the vacuum energy in the Hamiltonian
(\ref{HH1}), $E_{\rm vac}'$, differs from the vacuum energy
$E_{\rm vac}$ in the Hamiltonian (\ref{H2diag}) by terms coming
from the commutators of the operators $\hat{M}$ and
$\hat{M}^\dagger$.

\subsubsection{The case of the chiral pion}

Before we proceed with the diagonalisation of the Hamiltonian
(\ref{HH2}) in general form, let us consider the case of the
chiral pion as a paradigm for the general case without unnecessary
complications to cloud physics. The spin and spatial structure of
the pion
--- a $^1S_0$ state with $J=L=S=0$
--- can be chosen in matrix form so that the operator
$\hat{M}_{ss'}(\vec{p},\vec{p})$, creating the pion at rest, is
represented as
\be
\hat{M}_{ss'}(\vec{p},\vec{p})=\left[\kappa(\hat{\vec{p}})\right]_{ss'}\hat{M}(p),\quad
\kappa(\hat{\vec{p}})=\frac{i}{\sqrt{2}}\sigma_2Y_{00}(\hat{\vec{p}}).
\label{po}
\ee
Substitution of Eq.~(\ref{po}) into the Hamiltonian (\ref{HH2})
gives a simpler expression in terms of the radial operators
$\hat{M}(p)$ and $\hat{M}^\dagger(p)$,
$$
\hat{\cal H}_\pi=\int\frac{p^2d
p}{(2\pi)^3}2E_p\hat{M}^\dagger(p)\hat{M}(p)-\frac12\int\frac{p^2
dp}{(2\pi)^3}\frac{q^2dq}{(2\pi)^3}
\left\{T^{++}_\pi(p,q)\hat{M}^\dagger(p)M(q)\right.
$$
\be 
\left.+T^{+-}_\pi(p,q)\hat{M}^\dagger(q)\hat{M}^\dagger(p)
+T^{-+}_\pi(p,q)\hat{M}(p)\hat{M}(q)+T^{--}_\pi(p,q)
\hat{M}^\dagger(q)\hat{M}(p)\right\},
\label{HHpi}
\ee
where the amplitudes $T_\pi^{\pm\pm}(p,q)$ are trivially
built from the amplitudes $v^{\pm\pm}(\vec{p},\vec{q})$ (see
Eq.~(\ref{ampls})), after inclusion of the potential
$V(\vec{p}-\vec{q})$, integration over angular variables, and
taking a spin trace against the pion spin wave function $\kappa$,
as it was defined in Eq.~(\ref{po}).

To have a fully diagonalised Hamiltonian $\hat{\cal H}_\pi$ it is
now sufficient to perform a Bogoliubov transformation,
\be
\left\{
\begin{array}{l}
\hat{M}(p)=\hat{m}_\pi\vp_\pi^+(p)+\hat{m}_\pi^\dagger\vp_\pi^-(p)\\
\hat{M}^\dagger(p)=\hat{m}^\dagger_\pi\vp_\pi^+(p)+\hat{m}_\pi\vp_\pi^-(p),
\end{array}
\right.
\label{Mmpi}
\ee
or, inversely,
\ftnote{2}{Although Eqs.~(\ref{mMpi}) seem not to follow from
Eqs.~(\ref{Mmpi}) directly, which is a consequence of the naive
truncation we have made leaving only the pion instead of the whole
tower of mesonic states, in fact, Eqs.~(\ref{Mmpi}) and
(\ref{mMpi}) are self-consistent with each other at the given
level of truncation. See Eqs.~(\ref{Mmgen}) and (\ref{mMgen})
below.}
\be 
\left\{
\begin{array}{l}
\hat{m}_\pi=\ds\int\frac{p^2dp}{(2\pi)^3}\left[\hat{M}(p)\vp_\pi^+(p)-
\hat{M}^\dagger(p)\vp_\pi^-(p)\right]\\[2mm]
\hat{m}_\pi^\dagger=\ds\int\frac{p^2dp}{(2\pi)^3}\left[\hat{M}^\dagger
(p)\vp_\pi^+(p)- \hat{M}(p)\vp_\pi^-(p)\right],
\end{array}
\right.
\label{mMpi}
\ee
where the operators $\hat{m}_\pi^\dagger$ and $\hat{m}_\pi$ create and annihilate pions at rest.

Using the commutator of the radial operators,
\be
[\hat{M}(p),\;\hat{M}^\dagger
(q)]=\frac{(2\pi)^3}{p^2}\delta(p-q),
\ee
following from Eq.~(\ref{MMcom}), we find that
\be
[\hat{m}_\pi ,\; \hat{m}_\pi^\dagger]=
\int\frac{p^2dp}{(2\pi)^3}\left[\vp_\pi^{+2}(p)-\vp_\pi^{-2}(p)\right].
\ee
Therefore, requiring the canonical commutator between the pion
creation and annihilation operators to be $[m_\pi,m_\pi^\dagger]=1$, 
ensures that the  Bogoliubov amplitudes
$\vp_\pi^{\pm}$, playing the role of the pion wave functions, must
in turn be subject to the normalisation
\ftnote{3}{Notice that the wave functions $\vp^\pm_\pi$ can be
chosen real. This holds for all observable mesons as well. If the
mesonic wave function still contains an imaginary part then
Eqs.~(\ref{Mmpi})-(\ref{norm1}) should be changed accordingly, so
that, for example, $\vp_\pi^{\pm 2}$ in the normalisation
condition (\ref{norm1}) should be changed for $|\vp_\pi^{\pm}|^2$.
Below we discuss the situation with the tachyon where one is
forced to deal with an imaginary mass and, as a consequence, with
its imaginary wave functions.}
\be
\int\frac{p^2dp}{(2\pi)^3}\left[\vp_\pi^{+2}(p)-\vp_\pi^{-2}(p)\right]=1.
\label{norm1}
\ee

To complete the program of bosonic diagonalisation of the
Hamiltonian (\ref{H}), we need only to look for a pair of
$\vp_\pi^\pm (p)$ to diagonalise the hadronic Fock subspace
spanned by zero momentum pions, $|\pi_{P=0}\rangle =
m^\dagger_\pi|\Omega\rangle$, $|\Omega\rangle$ being the mesonic
vacuum (see the discussion below).

Now, if $\vp_\pi^+(p)$ and $\vp_\pi^-(p)$ are made to obey the following
set of equations,
\be
\left\{
\begin{array}{l}
[2E_p-M_\pi]\vp_\pi^+(p)=\ds\int\frac{\ds q^2dq}{\ds (2\pi)^3}
[T^{++}_\pi(p,q)\vp_\pi^+(q)+T^{+-}_\pi(p,q)\vp_\pi^-(q)]\\[0cm]
[2E_p+M_\pi]\vp_\pi^-(p)=\ds\int\frac{\ds q^2dq}{\ds (2\pi)^3}
[T^{-+}_\pi(p,q)\vp_\pi^+(q)+T^{--}_\pi(p,q)\vp_\pi^-(q)],
\end{array}
\right.
\label{bsp}
\ee
we can, by a simple substitution in Eq.~(\ref{HHpi}), get rid of
the anomalous Bogoliubov terms to have $\langle\Omega|\hat{\cal
H}_\pi|\pi_{P=0}\pi_{P=0}\rangle =0$ and $\langle
\pi_{P=0}\pi_{P=0}|\hat{\cal H}_\pi|\Omega\rangle=0$. The
non-anomalous terms can be readily summed to yield:
\be
\hat{\cal H}_\pi=M_\pi \hat{m}_\pi^\dagger
\hat{m}_\pi+\ldots,\quad \langle \pi_{P=0}|\hat{\cal
H}_\pi|\pi_{P=0}\rangle = M_\pi,
\label{diagtot}
\ee
where the ellipsis denotes terms suppressed by $N_C$. For example,
the quartic term in the pion creation/annihilation operators
responsible for the pion-pion scattering can be identified among
the latter.

Notice the difference between the quark BCS vacuum $|0\rangle$,
annihilated by the dressed quark operators $\hat{b}$ and
$\hat{d}$, and the mesonic vacuum $|\Omega\rangle$, annihilated by
the mesonic operators, for example, by $\hat{m}_\pi$. The two
vacua are related by a unitary transformation,
$|0\rangle=U^\dagger|\Omega\rangle$, with the operator $U^\dagger$ creating pairs
of mesons and taking such a form that, for example,
$$
\hat{m}_\pi|\Omega\rangle=\hat{m}_\pi U^\dagger|0\rangle
=U^\dagger(U\hat{m}_\pi U^\dagger)|0\rangle \propto
U^\dagger\hat{M}(p)|0\rangle=0.
$$
Since creation of any extra quark-antiquark pair is suppressed in
the large-$N_C$ limit, so is  the deviation of the operator $U^\dagger$
from unity. As a result, the chiral condensate calculated at
BCS level, using the BCS vacuum $|0\rangle$, coincides, up to
$1/N_C$ corrections, with the full quark condensate calculated in
the mesonic vacuum $|\Omega\rangle$.

It is easy to recognise in Eq.~(\ref{bsp}) the Bethe--Salpeter equation for the
pion studied in a number papers
\cite{Orsay2,Lisbon}. To see this equivalence we proceed in five steps.

{\noindent Step 1:}

First let us define dressed quarks and derive the mass-gap
equation. To this end we consider the quark mass operator in the
rainbow approximation,
\be
i\Sigma(\vec{p})=\int\frac{d^4k}{(2\pi)^4}V(\vec{p}-\vec{k})\Gamma
S_F(\vec{k},k_0)\Gamma, 
\label{Sigma01} 
\ee 
where $S_F({\vec p},p_0)$, 
\be 
S_F({\vec p},p_0)=\frac{\Lambda^{+}({\vec
p})\gamma_0}{p_0-E_p+i\epsilon}+ \frac{{\Lambda^{-}}({\vec
p})\gamma_0}{p_0+E_p-i\epsilon},
\label{Feynman}
\ee
$$
\Lambda^\pm(\vec{p})=\frac12[1\pm\gamma_0\sin\vp_p\pm(\vec{\alpha}\hat{\vec{p}})\cos\vp_p],
$$
is the Feynman propagator of the dressed quark with the dispersive
law $E_p$. $\Gamma$ is a generic Dirac matrix defining the Lorentz
nature of the confining interaction. Parameterising
$\Sigma(\vec{p})$ and $E_p$ as
\be
\Sigma(\vec{p})=[A_p-m]+(\vec{\gamma}\hat{\vec{p}})[B_p-p],\quad
E_p=A_p\sin\vp_p+B_p\cos\vp_p,
\ee
we find that
\be
S_F^{-1}(p_0,\vec{p})=\gamma_0p_0-(\vec{\gamma}\hat{\vec{p}})B_p-A_p.
\ee
Taking the integral in the energy in Eq.~(\ref{Sigma01}), we
arrive at the self-consistency condition,
\be
A_p\cos\vp_p-B_p\sin\vp_p=0,
\label{mge2}
\ee
which is the mass-gap equation for the chiral angle $\vp_p$. For
$\Gamma=\gamma_0$ the auxiliary functions $A_p$ and $B_p$ are
given by Eqs.~(\ref{A}) and (\ref{B}). Notice, however, that the
general form of Eq.~(\ref{mge2}) holds for any Lorentz
nature of confinement (see \cite{Lisbon} for a detailed
discussion). This completes the matching, at BCS level,
between the Hamiltonian and diagrammatic approaches to the theory.

{\noindent Step 2:}

Proceed beyond BCS level and consider the Salpeter equation
(in the Dirac representation) for a generic meson in the rest
frame,
\be
\chi({\vec p};M)=-i\int\frac{d^4q}{(2\pi)^4}V(\vec{p}-\vec{q})\;
\Gamma S_F({\vec q},q_0+M/2)\chi({\vec q};M)S_F({\vec q},q_0-M/2)\Gamma,
\label{GenericSal}
\ee
where $\chi({\vec p};M)$ stands for the mesonic Salpeter amplitude.
\bigskip

{\noindent Step 3:}

Due to the instantaneous nature of $V(\vec{p}-\vec{q})$, it is a
simple matter of poles logistics to see that, upon integration in
the energy,  the only surviving combinations of the Feynman
projectors $\Lambda^{\pm}$ entering in Eq.~(\ref{GenericSal}) are
$(\Lambda^+\gamma_0)\chi(\Lambda^-\gamma_0)$ and
$(\Lambda^-\gamma_0)\chi(\Lambda^+\gamma_0)$, so that we can
decompose $\chi$ in two distinct amplitudes, $\chi^{[+]}$ and
$\chi^{[-]}$. To this end we get rid of the energy denominators in
Eq.~(\ref{GenericSal}) by taking the integral in the energy, 
\be
\int_{-\infty}^{\infty}\frac{dq_0}{2\pi i}\left[\frac{1}{q_0\pm
M/2-E_q+i\epsilon}\right] \left[\frac{1}{q_0\mp
M/2+E_q-i\epsilon}\right]=-\frac{1}{2E_q\mp M}, 
\ee 
and define:
$$
\chi^{[+]}(\vec{q};M)=\frac{\chi(\vec{q};M)}{2E_q-M},\quad
\chi^{[-]}(\vec{q};M)=\frac{\chi(\vec{q};M)}{2E_q+M}.
$$

Then the Bethe--Salpeter equation
(\ref{GenericSal}) amounts to a system of two coupled equations:
\be
\left\{
\begin{array}{l}
[2E_p-M]\chi^{[+]}=-\ds\int\frac{d^3q}{(2\pi)^3}V(\vec{p}-\vec{q})\;
\Gamma\left[(\Lambda^+\gamma_0)\chi^{[+]}(\Lambda^-\gamma_0)
+(\Lambda^-\gamma_0)\chi^{[-]}(\Lambda^+\gamma_0)\right]\Gamma\\[0mm]
[2E_p+M]\chi^{[-]}=-\ds\int\frac{d^3q}{(2\pi)^3}V(\vec{p}-\vec{q})\;
\Gamma\left[(\Lambda^+\gamma_0)\chi^{[+]}(\Lambda^-\gamma_0)
+(\Lambda^-\gamma_0)\chi^{[-]}(\Lambda^+\gamma_0)\right]\Gamma.
\end{array}
\label{Salpeterlev3}
\right.
\ee
\bigskip

{\noindent Step 4:}

Sandwich both equations in (\ref{Salpeterlev3}) between the quark
spinors and use the definition of the projectors,
\be
\Lambda^+(\vec{p})=\sum_{s_1s_2}u_{s_1}({\vec p})\otimes
u^\dagger_{s_2} ({\vec p}),\quad
\Lambda^-(\vec{p})=\sum_{s_1s_2}v_{s_1}(-{\vec p})\otimes
v^\dagger_{s_2} (-{\vec p}),
\label{Lambdas}
\ee
to cast Eq.~(\ref{Salpeterlev3}) as
\be
\left\{
\begin{array}{r}
[2E_p-M]\left[\bar{u}_{s_1}\chi^{[+]}v_{s_2}\right] =-\ds\sum_{s_3s_4}\int\frac{d^3q}{(2\pi)^3}V(\vec{p}-\vec{q})
\left\{[\bar{u}_{s_1}\Gamma u_{s_3}][\bar{u}_{s_3}\chi^{[+]}v_{s_4}]
[\bar{v}_{s_4}\Gamma v_{s_2}]\right.\\
\left.+[\bar{u}_{s_1}\Gamma v_{s_3}]
[\bar{v}_{s_3}\chi^{[-]}u_{s_4}][\bar{u}_{s_4}\Gamma v_{s_2}]\right\}\hphantom{.}\\[3mm]
[2E_p+M]\left[\bar{v}_{s_1}\chi^{[-]}u_{s_2}\right] =-\ds\sum_{s_3s_4}\int\frac{d^3q}{(2\pi)^3}V(\vec{p}-\vec{q})
\left\{[\bar{v}_{s_1}\Gamma u_{s_3}][\bar{u}_{s_3}\chi^{[+]}v_{s_4}]
[\bar{v}_{s_4}\Gamma u_{s_2}]\right.\\
\left.+[\bar{v}_{s_1}\Gamma v_{s_3}]
[\bar{v}_{s_3}\chi^{[-]}u_{s_4}][\bar{u}_{s_4}\Gamma u_{s_2}]\right\}.
\label{Salpeterlev4}
\end{array}
\right.
\ee
\bigskip

{\noindent Step 5:}

Define $\Phi_{s_1s_2}^+=[\bar{u}_{s_1}\chi^{[+]}v_{s_2}]$ and,
similarly, $\Phi_{s_1s_2}^-=[\bar{v}_{s_1}\chi^{[-]}u_{s_2}]$ to get: 
\be 
\left\{
\begin{array}{l}
[2E_p-M]\Phi_{s_1s_2}^+=-\ds\sum_{s_3s_4}\int\frac{d^3q}{(2\pi)^3}V(\vec{p}-\vec{q})
\left\{[v^{++}]_{s_1s_2s_3s_4}\Phi_{s_3s_4}^+ +[v^{+-}]_{s_1s_2s_3s_4}\Phi_{s_3s_4}^-\right\}\\[1mm]
[2E_p+M]\Phi_{s_1s_2}^-=-\ds\sum_{s_3s_4}\int\frac{d^3q}{(2\pi)^3}V(\vec{p}-\vec{q})
\left\{[v^{-+}]_{s_1s_2s_3s_4}\Phi_{s_3s_4}^+ +[v^{--}]_{s_1s_2s_3s_4}\Phi_{s_3s_4}^-\right\}.
\label{Salpeterlev5}
\end{array}
\right.
\ee

In the coefficients $v^{\pm\pm}$ one can easily recognise the amplitudes
which appeared in the Hamiltonian (\ref{HH2}). For the case
$\Gamma=\gamma_0$ they are given in Eq.~(\ref{ampls}).
Further details and explicit forms for the amplitudes (\ref{ampls}) in terms of the
chiral angle can be found in refs.~\cite{Lisbon}.

For establishing graphical rules it is convenient to include the
potential into the definition of the amplitudes and to rewrite
them in the form:
\be
\begin{array}{l}
[T^{++}(\vec{p},\vec{q})]_{s_1s_2s_3s_4}=[\bar{u}_{s_1}(\vec{p})\Gamma u_{s_3}(\vec{q})]
[-V(\vec{p}-\vec{q})][\bar{v}_{s_4}(-\vec{q})\Gamma v_{s_2}(-\vec{p})],\\[0cm]
[T^{+-}(\vec{p},\vec{q})]_{s_1s_2s_3s_4}=[\bar{u}_{s_1}(\vec{p})\Gamma v_{s_3}(-\vec{q})]
[-V(\vec{p}-\vec{q})][\bar{u}_{s_4}(\vec{q})\Gamma v_{s_2}(-\vec{p})],\\[0cm]
[T^{-+}(\vec{p},\vec{q})]_{s_1s_2s_3s_4}=[\bar{v}_{s_1}(-\vec{p})\Gamma u_{s_3}(\vec{q})]
[-V(\vec{p}-\vec{q})][\bar{v}_{s_4}(-\vec{q})\Gamma v_{s_2}(\vec{p})],\\[0cm]
[T^{--}(\vec{p},\vec{q})]_{s_1s_2s_3s_4}=[\bar{v}_{s_1}(-\vec{q})\Gamma v_{s_3}(-\vec{p})]
[-V(\vec{p}-\vec{q})][\bar{u}_{s_4}(\vec{p})\Gamma u_{s_2}(\vec{q})],
\end{array}
\label{ampls2}
\ee
or simply,
\be
\begin{array}{c}
T^{++}=[\bar{u}\Gamma u][-V][\bar{v}\Gamma v],\quad T^{+-}=[\bar{u}\Gamma v][-V][\bar{u}\Gamma v],\\
T^{-+}=[\bar{v}\Gamma u][-V][\bar{v}\Gamma u],\quad T^{--}=[\bar{v}\Gamma v][-V][\bar{u}\Gamma u].
\end{array}
\ee

In Fig.~1 we depict the steps leading to Eq.~(\ref{Salpeterlev5})
with the amplitudes (\ref{ampls2}). In this figure, the first and
the second lines represent the r.h.s. of the first equation in the
system (\ref{Salpeterlev3}), sandwiched between the quark spinors
$\bar{u}$ and $v$, and the r.h.s. of the first equation in the
system (\ref{Salpeterlev5}), respectively. Similar diagrams can be
drawn for the second equation of both systems --- the only
difference will be the ordering of the spinors $u$ and $v$.
Thus we introduce a diagrammatic technique relating Bethe--Salpeter
amplitudes and vertices to the coefficients of the Hamiltonian in the
representation of mesonic operators.

\begin{figure}[t]
\begin{center}
\includegraphics[width=14.5 cm]{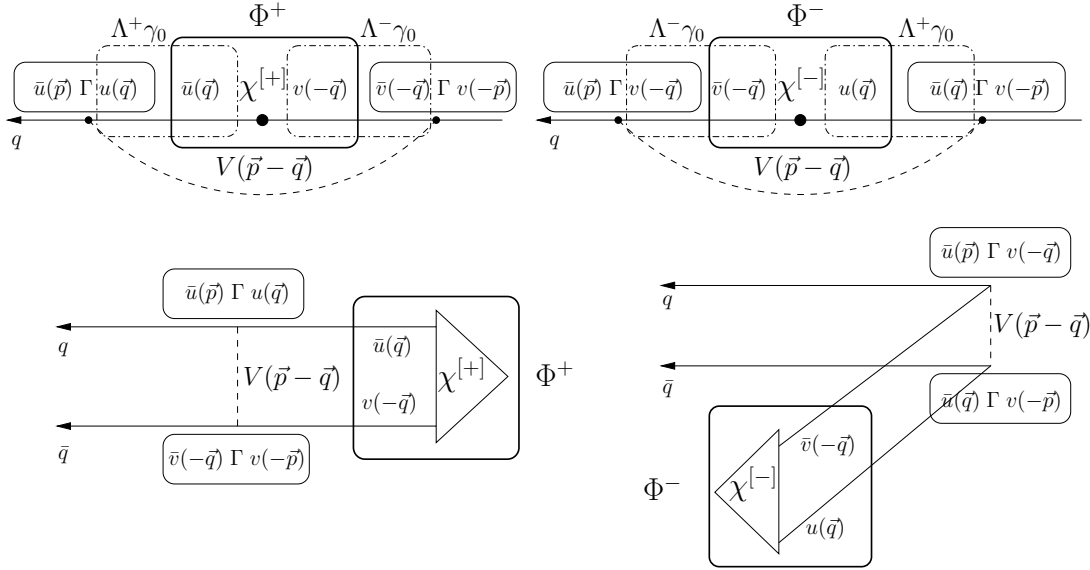}
\end{center}
\caption{Going from the Dirac representation to the spin
representation: one uses the definitions of the projectors $\Lambda^{\pm}$,
Eq.~(\ref{Lambdas}),
(dash-dotted line) to reshuffle the spinors towards the $\Gamma$ vertices
(solid line). The
Salpeter amplitudes $\chi^{[\pm]}$ are vectorially multiplied
by the adjacent spinors in order to construct two functions --- two-by-two matrices in spins: $\Phi^+$ and
$\Phi^-$ (fat solid line).}
\end{figure}

Now we need to project the spin-angular wave functions
$\Phi_{s_1s_2}^\pm(\vec{p})$ onto the wave function with the
definite total momentum, parity, and charge conjugation number. For
the pion this is quite easily done. We have
\be
\Phi_{s_1s_2}^{\pm}(\vec{p})=\left[\frac{i}{\sqrt{2}}\sigma_2\right]_{s_1s_2}
Y_{00}(\hat{\vec{p}})\vp_\pi^\pm(p),\quad
Y_{00}(\hat{\vec{p}})=\frac{1}{\sqrt{4\pi}},
\label{vppi}
\ee
which should be compared with Eq.~(\ref{po}). Finally, we need
only to perform a spin trace and to introduce the $T_\pi^{\pm\pm}$
amplitudes for the pion, such as,
\be
T_\pi^{++}(p,q)=-\frac{1}{2}\int d\Omega_p
d\Omega_qY_{00}^*(\hat{\vec{p}})V(\vec{p}-\vec{q})
Y_{00}(\hat{\vec{q}})Sp\left\{\sigma_2[\bar{u}(\vec{p})\Gamma
u(\vec{q})]\sigma_2 [\bar{v}(-\vec{q})\Gamma v(-\vec{p})]\right\},
\label{Tpp}
\ee
to arrive at the pion Bethe--Salpeter equation in the spin
representation, which coincides with the bosonic mass-gap equation
(\ref{bsp}). For the case of the harmonic oscillator potential it
can be rewritten in a matrix form,
\be
\left[ \left[-K_0^3\frac{d^2}{dp^2}+2E_p\right]\left[
\begin{array}{cc}
1&0\\0&1
\end{array}
\right]
+K_0^3
\left[
\frac{\vp^{'2}_p}{2}+\frac{\cos^2\vp_p}{p^2}\right]\left[
\begin{array}{cc}1&1\\1&1\end{array}
\right]
-
M_\pi\left[
\begin{array}{cc}1&0\\0&-1\end{array}
\right]
\right]
\left[
\begin{array}{c}
\nu_\pi^+(p)\\
\nu_\pi^-(p)
\end{array}
\right]=0, \label{hop}
\ee
for the radial wave functions $\nu^\pm_\pi(p)=p\vp^\pm_\pi(p)$
which are rescaled so as to obey the one-dimensional normalisation
\cite{Lisbon},
\be
\int dp[\nu^{+2}_\pi(p)-\nu^{-2}_\pi(p)]=1.
\ee
Comparison of the Hamiltonian (\ref{HHpi}) with the bound-state equation
(\ref{Salpeterlev5}) allows us to establish a dictionary between the terms in
(\ref{HHpi}) and the diagrammatic rules used in the literature in order
to arrive at the Bethe--Salpeter equation (\ref{Salpeterlev5}). These rules are
easily deduced from Fig.~1 and will be exploited in what follows in order to
diagonalise the full Hamiltonian (\ref{HH2}).

Notice several important properties of the bound-state equation
(\ref{bsp}) and its solutions. In the chiral limit the pion mass
vanishes, $M_\pi=0$, and the pion wave functions are real and are
related to one another and to the chiral angle as
$\vp_\pi^+(p)=-\vp_\pi^-(p)={\cal N}_\pi\sin\vp_p$, ${\cal N}_\pi$
being the norm. Let us use the simple case of the harmonic
oscillator kernel to illustrate this. Using the explicit form of
the $T$-amplitudes taken from the papers \cite{Lisbon}, one easily
finds that, for the case of the $^1S_0$ state,
$T^{++}_\pi(p,q)-T^{+-}_\pi(p,q)=T^{-+}_\pi(p,q)-T^{--}_\pi(p,q)=
\frac{(2\pi)^3}{p^2}\delta(p-q)K_0^3\Delta_{q}$, where the
Laplacian is actually reduced to its radial part (see
Eq.~(\ref{hop})). Substituting this result in the bound-state
equation (\ref{bsp}) for $\vp_\pi^+(p)=-\vp_\pi^-(p)\equiv\vp_\pi$,
and using $M_\pi=0$, we get
\be
[-K_0^3\Delta_p+2E_p]\vp_\pi=0,
\ee
which coincides with the mass-gap equation (\ref{mgho}) with
$\vp_\pi=\psi=\sin\vp_p$. Derivation of the pion bound-state
equation as a Shr{\" o}dinger-like equation for the generic form
of the potential is given in Appendix A.

Thus we arrive at a very important conclusion that, in the chiral
limit, {\em the mass-gap equation is the bound-state equation for
the Goldstone boson at rest, responsible for the spontaneous
breaking of chiral symmetry. Since the chiral pion is this
Goldstone boson, then we are forced to see it already at the BCS
level and it will remain being the massless Goldstone boson even
when we go beyond BCS.} The properties of the pion have been
widely discussed in the literature--- see, for example,
\cite{Orsay,Orsay2,Lisbon} for QCD in 3+1 and \cite{tHooft,2d} for
QCD in 1+1.

Beyond the chiral limit the solution to the bound-state equation (\ref{bsp})
takes an approximate form:
\be
\vp_\pi^\pm(p)=\tilde{\cal N}_\pi\left[\pm\frac{1}{\sqrt{M_\pi}}\sin\vp_p+
\sqrt{M_\pi}\Delta_p\right],\quad
\tilde{\cal N}_\pi^2=4\int_0^\infty\frac{p^2dp}{(2\pi)^3}\Delta_p\sin\vp_p,
\label{vppm}
\ee
where all corrections of higher order in the pion mass are neglected and the function
$\Delta_p$ obeys a reduced $M_\pi$--independent equation
(see, for example, \cite{Lisbon} or the papers \cite{2d} where such an
equation for $\Delta_p$ is discussed in
two-dimensional QCD).

Notice that, although the dressed quark dispersive law and $T$-amplitudes for the pion are real
(see Appendix A for the details),
the bound-state equation (\ref{bsp}) admits a tachyonic solution with $M_{\rm tach}^2<0$
(see also the
papers \cite{Orsay2} where a tachyonic solution was found for the bound-state  equation
in the trivial
vacuum). For $M_{\rm tach}=i|M_{\rm tach}|$, the tachyon wave functions fail to be real and
acquire imaginary parts, in accordance with Eq.~(\ref{vppm}). It is easy to verify that they
obey a simple transformation rule, $(\vp^\pm_{\rm tach})^*\propto \vp^\mp_{\rm tach}$.

\subsubsection{The case of the $\rho$-meson}

Compared to the pion, the case of the $\rho$-meson involves an
extra complication, a consequence of its richer spatial structure.
The Hamiltonian (\ref{HH2}) does not commute with the operators of
the total spin $\vec{S}$ and the angular momentum $\vec{L}$, so it
cannot be diagonalised in terms of the $^{2S+1}L_J$ states. The
appropriate basis is given by the set of physical observable
mesons $J^{PC}$. Then, although the lowest $0^{-+}$ state --- the
chiral pion --- is represented by just only one term, a pure
$^1S_0$ state, higher states are in general a mixture of angular
momentum states, like, for instance, the case of $1^{--}$
$\rho$-meson, which is a mixture of $^3S_1$ and $^3D_1$ terms.
This problem does not appear in two-dimensional QCD and is a
consequence of a richer structure of the spatial dimensions in the
four-dimensional theory. Therefore, in order to formulate the
eigenvalue problem for the $\rho$-meson, one should introduce a
set of four wave functions,
$\left\{\vp^{\pm}_{^3S_1}\equiv\vp^\pm_0\right.$,
$\left.\vp^{\pm}_{^3D_1}\equiv\vp^\pm_2\right\}$,
and write the system of four coupled equation,
\be
\left\{
\begin{array}{l}
[2E_p-M_\rho]\vp_0^+(p)=\ds\int\frac{\ds q^2dq}{\ds (2\pi)^3}
[T^{++}_{00}\vp_0^+(q)+T^{++}_{02}\vp_2^+(q)+T^{+-}_{00}\vp_0^-(q)+T^{+-}_{
02}\vp_2^-(q)]\\[0cm]
[2E_p+M_\rho]\vp_0^-(p)=\ds\int\frac{\ds q^2dq}{\ds (2\pi)^3}
[T^{-+}_{00}\vp_0^+(q)+T^{-+}_{02}\vp_2^+(q)+T^{--}_{00}\vp_0^-(q)+T^{--}_{
02}\vp_2^-(q)]\\[0cm]
[2E_p-M_\rho]\vp_2^+(p)=\ds\int\frac{\ds q^2dq}{\ds (2\pi)^3}
[T^{++}_{22}\vp_2^+(q)+T^{++}_{20}\vp_0^+(q)+T^{+-}_{22}\vp_2^-(q)+T^{+-}_{
20}\vp_0^-(q)]\\[0cm]
[2E_p+M_\rho]\vp_2^-(p)=\ds\int\frac{\ds q^2dq}{\ds (2\pi)^3}
[T^{-+}_{22}\vp_2^+(q)+T^{-+}_{20}\vp_2^+(q)+T^{--}_{22}\vp_2^-(q)+T^{--}_{
20}\vp_2^-(q)],
\end{array}
\right.
\label{bsrho}
\ee
where the $T$-amplitudes can be built explicitly
using the graphical rules established above and with the help of
the spin-angular wave functions
$[\kappa_{^3S_1}(\hat{\vec{p}})]_{s_1s_2}\equiv[\kappa_0(\hat{\vec{p}})]_{s_1s_2}$
and
$[\kappa_{^3D_1}(\hat{\vec{p}})[_{s_1s_2}\equiv[\kappa_2(\hat{\vec{p}})]_{s_1s_2}$.
For example, similarly to Eq.~(\ref{Tpp}) we have,
\be
T_{02}^{++}(p,q)=-\int d\Omega_p d\Omega_qV(\vec{p}-\vec{q})
Sp\left\{\kappa_0(\hat{\vec{p}})[\bar{u}(\vec{p})\Gamma
u(\vec{q})]\kappa_2(\hat{\vec{q}}) [\bar{v}(-\vec{q})\Gamma
v(-\vec{p})]\right\},
\label{Tpp2}
\ee
and so on.

As an example, we
give here the explicit form of the Bethe--Salpeter equation (\ref{bsrho})
for the harmonic oscillator potential \cite{Lisbon}:
$$
\left[
\left[-\frac{d^2}{dp^2}+2E_p\right]
\left[
\begin{array}{cc}1&0\\0&1\end{array}
\right]
+
\frac{6}{p^2}\left[\begin{array}{cc}1&0\\0&0\end{array}\right]
+
\frac{\vp^{'2}_p}{2}\left[\begin{array}{cc}3-\sigma_1&
-2\sqrt{2}\\-2\sqrt{2}&3+\sigma_1\end{array}\right]\right.
+
\frac{2(1-\sin\vp_p)}{3p^2}\left[\begin{array}{cc}-4&
\sqrt{2}\\
\sqrt{2}&-4\end{array}\right]
$$
\be
\left.
-
\frac{\cos^2\vp_p}{3p^2}\left[\begin{array}{cc}2(1-\sigma_1)&
\sqrt{2}(1-\sigma_1)\\\sqrt{2}(1-\sigma_1)&1-\sigma_1\end{array}\right]
-
M_\rho\left[\begin{array}{cc}\sigma_3&0\\0&-\sigma_3\end{array}\right]
\right]
\left[
\begin{array}{c}
\nu_2^+(p)\\
\nu_2^-(p)\\
\nu_0^+(p)\\
\nu_0^-(p)
\end{array}
\right]=0,
\label{horho}
\ee
where $\sigma$'s are the Pauli matrices and, similarly to the case of
the pion, the new radial wave functions are defined as
$\nu^\pm_{0,2}(p)=p\vp^\pm_{0,2}(p)$.

Naively, one may conclude, from the bound-state equation
(\ref{bsrho}), that the $\rho$-meson has to be described by a
four-component wave function, rather than by just a two-component
wave function, as it was the case with the pion. This conclusion
is erroneous, since the doubling of the number of equations in
(\ref{bsrho}) is a consequence of the usage of an inappropriate
basis $^{2S+1}L_J$. Indeed, together with the light $\rho$-meson,
we must have a second solution for the the bound-state equation
(\ref{bsrho}) needed to describe the heavier vectorial partner of
the $\rho$. The easiest way to show this is to neglect
off-diagonal terms in Eq.~(\ref{bsrho}), thus splitting the system
of four equation into two independent systems of two equations
each, for $\vp_0^\pm$ and $\vp_2^\pm$, respectively. The
eigenvalues found for these two independent eigenvalue problems
give the masses of the pure $^3S_1$ and $^3D_1$ states, which
should be mixed then as 
\be 
{\rm det}\left(
\begin{array}{cc}
M_{^3S_1}-M_\rho&\Delta_{SD}\\
\Delta_{DS}&M_{^3D_1}-M_\rho
\end{array}
\right)=0, 
\label{MSD} 
\ee 
with $\Delta_{SD}$ and $\Delta_{DS}$
being the contributions of the restored off-diagonal terms of the
full Hamiltonian (\ref{HH2}). The lighter solution represents the
physical $\rho$-meson in the basis $J^{PC}$.
In other words, if an appropriate basis --- which
diagonalises the three-dimensional spatial part of the Hamiltonian
(\ref{HH2})
--- is chosen from the very beginning, each physical mesonic state
is fully described by just a two-component wave function and the
corresponding eigenvalue --- the physical mass of the meson.
Notice that the transformation (\ref{MSD}) can be viewed as yet
another Bogoliubov-Valatin transformation parameterised by the
$S$-wave--$D$-wave mixing angle.

Now, with the $\rho$-meson included, the diagonalised Hamiltonian
takes an obvious form,
\be
\hat{\cal H}=M_\pi m_\pi^\dagger m_\pi+M_\rho m_\rho^\dagger
m_\rho+ M_{\rho'} m_{\rho'}^\dagger m_{\rho'}\ldots,
\ee
where $\rho'$ denotes the heavier partner of $\rho$ which is also
a solution of Eq.~(\ref{MSD}) and the operators
creating/annihilating the vector mesons are defined similarly to
the pionic case, Eqs.~(\ref{po}) and (\ref{Mmpi}), through the
$1^{--}$ wave functions $\vp^\pm_{\rho,\rho'}$. We are now ready
to consider the general case.

\subsubsection{The general case}

Now we return to the general case of the Hamiltonian (\ref{HH2})
and diagonalise it in terms of the mesonic compound operators. We work
in the basis $\{n,J^{PC}\}$; $n$ being the radial quantum number.
We also assume that this basis includes all states with the given
$J^{PC}$, so that the set of such states is complete and has a
one-to-one correspondence with the complete set $^{2S+1}L_J$. For
the sake of simplicity, we denote each mesonic state as
$\{n,\nu\}$, where the set of quantum numbers $\nu$ is not only
based on the $J^{PC}$ classification scheme, but also identifies
each meson in a given $nJ^{PC}$ multiplet, as it was discussed in
the case of $\rho$ and $\rho'$. Then, for a given pair
$\{n,\nu\}$, representing a physical meson, we introduce a
two-component radial wave function $\vp^\pm_{n\nu}(p)$ and the
spin-angular wave function $\kappa_\nu(\hat{\vec{p}})$, as it was
done in the pion case --- see Eq.~(\ref{po}). The operators
$\hat{M}^\dagger$ and $\hat{M}$ of Eq.(\ref{operatorsMM}),
entering the Hamiltonian (\ref{HH2}), can be expanded using this
basis (Eq.~(\ref{po}) for the pion follows from this relation if
we take only the lowest $\nu=0^{-+}$),
\be
\hat{M}_{ss'}(\vec{p},\vec{p})=
\sum_\nu[\kappa_\nu(\hat{\vec{p}})]_{ss'}\hat{M}_\nu(p).
\ee
Now we use the completeness property of the set
$\{\kappa_\nu(\hat{\vec{p}})\}$ together with the fact that this
set diagonalises the spin and the angular structure of the
Hamiltonian (\ref{HH2}), to find:
\be
\sum_{s_1s_2}\int
d\Omega_p[\kappa_\nu(\hat{\vec{p}})]_{s_1s_2}^*
[\kappa_{\nu'}(\hat{\vec{p}})]_{s_2s_1}=\delta_{\nu\nu'},
\label{59}
\ee
and
\be
\begin{array}{c}
\ds \sum_{s_1s_2s_3s_4}\int d\Omega_pd\Omega_qV(\vec{p}-\vec{q})
[v^{++}(\vec{p},\vec{q})]_{s_1s_3s_4s_2}[\kappa_\nu(\hat{\vec{p}})]^*_{s_2s_1
}[\kappa_{\nu'}(\hat{\vec{q}})]_{s_4s_3}=
-T^{++}_\nu(p,q)\delta_{\nu\nu'},\\
\ds \sum_{s_1s_2s_3s_4}\int d\Omega_pd\Omega_qV(\vec{p}-\vec{q})
[v^{+-}(\vec{p},\vec{q})]_{s_1s_3s_4s_2}[\kappa_\nu(\hat{\vec{q}})]^*_{s_2s_1
}[\kappa_{\nu'}(\hat{\vec{p}})]_{s_3s_4}^*=
-T^{+-}_\nu(p,q)\delta_{\nu\nu'},\\
\ds \sum_{s_1s_2s_3s_4}\int d\Omega_pd\Omega_qV(\vec{p}-\vec{q})
[v^{-+}(\vec{p},\vec{q})]_{s_1s_3s_4s_2}[\kappa_\nu(\hat{\vec{p}})]_{s_1s_2}
[\kappa_{\nu'}(\hat{\vec{q}})]_{s_4s_3}=
-T^{-+}_\nu(p,q)\delta_{\nu\nu'},\\
\ds \sum_{s_1s_2s_3s_4}\int d\Omega_pd\Omega_qV(\vec{p}-\vec{q})
[v^{--}(\vec{p},\vec{q})]_{s_1s_3s_4s_2}[\kappa_\nu(\hat{\vec{q}})]_{s_1s_2}
[\kappa_{\nu'}(\hat{\vec{p}})]^*_{s_3s_4}=
-T^{--}_{\nu}(p,q)\delta_{\nu\nu'}.
\end{array}
\ee

As a result, the Hamiltonian (\ref{HH2}) takes the form, in terms
of the radial operators $\{\hat{M}_\nu^\dagger(p)\}$ and
$\{\hat{M}_\nu(p)\}$ as:
$$
\hat{\cal H}=\sum_\nu\int\frac{p^2d p}{(2\pi)^3}2E_p\hat{M}_\nu^\dagger(p)\hat{M}_\nu(p)
-\frac12\sum_\nu\int\frac{p^2dp}{(2\pi)^3}\frac{q^2dq}{(2\pi)^3}\left\{
T_\nu^{++}(p,q)\hat{M}_\nu^\dagger(p)\hat{M}_\nu(q)\right.
$$
\be
\left.+T^{+-}_\nu(p,q)\hat{M}_\nu^\dagger(q)\hat{M}_\nu^\dagger(p)
+T^{-+}_\nu(p,q)\hat{M}_\nu(p)\hat{M}_\nu(q)+T^{--}_\nu(p,q)
\hat{M}_\nu^\dagger(q)\hat{M}_\nu(p)\right\},
\label{HHgen}
\ee
where all radial excitations are now disentangled from one another.

Then the generalisation of the relations (\ref{Mmpi}),
(\ref{mMpi}) is trivial, 
\be 
\left\{
\begin{array}{l}
\hat{M}_\nu(p)=\sum_n[\hat{m}_{n\nu}\vp_{n\nu}^+(p)+\hat{m}_{n\nu}^\dagger\vp_{n\nu}^-(p)]\\
\hat{M}_\nu^\dagger(p)=\sum_n
[\hat{m}^\dagger_{n\nu}\vp_{n\nu}^+(p)+\hat{m}_{n\nu}\vp_{n\nu}^-(p)],
\end{array}
\right.
\label{Mmgen}
\ee
\be
\left\{
\begin{array}{l}
\hat{m}_{n\nu}=\ds\int\frac{p^2dp}{(2\pi)^3}\left[\hat{M}_\nu(p)\vp_{n\nu}^+(p)-
\hat{M}^\dagger_\nu(p)\vp_{n\nu}^-(p)\right]\\
\hat{m}_{n\nu}^\dagger=\ds\int\frac{p^2dp}{(2\pi)^3}\left[\hat{M}_\nu^\dagger
(p)\vp_{n\nu}^+(p)- \hat{M}_\nu(p)\vp_{n\nu}^-(p)\right].
\end{array}
\right.
\label{mMgen}
\ee

As in the pion case, we build the commutators of the operators
$\hat{m}_{n\nu}$ and $\hat{m}_{m\nu}^\dagger$,
\be
\begin{array}{l}
[\hat{m}_{n\nu},\;\hat{m}_{m\nu}^\dagger]=\ds
\int\frac{p^2dp}{(2\pi)^3}\left[\vp_{n\nu}^{+}(p)\vp_{m\nu}^{+}(p)-
\vp_{n\nu}^{-}(p)\vp_{m\nu}^{-}(p)\right],\\[3mm]
[\hat{m}_{n\nu},\;\hat{m}_{m\nu}]=\ds
\int\frac{p^2dp}{(2\pi)^3}\left[\vp_{m\nu}^{+}(p)\vp_{n\nu}^{-}(p)-
\vp_{m\nu}^{-}(p)\vp_{n\nu}^{+}(p)\right],
\end{array}
\ee
and require that they should obey the standard bosonic algebra,
that is, $[\hat{m}_{n\nu},\;\hat{m}_{m\nu}^\dagger]=\delta_{mn}$,
and $[\hat{m}_{n\nu},\;\hat{m}_{m\nu}]=0$. Therefore, similarly to
Eq.~(\ref{norm1}), for any given $\nu$, we arrive at the
normalisation and orthogonality conditions for $\vp$'s,
\ftnote{4}{Strictly speaking, the orthogonality condition for the
mesonic wave functions should also include their spin-angular part. For example,
$$
\sum_{s_1s_2}\int d\Omega_p[\kappa_\nu(\hat{\vec{p}})]^*_{s_1s_2}
[\kappa_{\nu'}(\hat{\vec{p}})]_{s_2s_1}
\int\frac{p^2dp}{(2\pi)^3}\left[\vp_{n\nu}^{+}(p)\vp_{m\nu'}^{+}(p)-
\vp_{n\nu}^{-}(p)\vp_{m\nu'}^{-}(p)\right]=\delta_{nm}\delta_{\nu\nu'},
$$
where, for non-coinciding $\nu$'s, the l.h.s. vanishes due to the angular
integration --- see the orthogonality condition for $\kappa$'s,
Eq.~(\ref{59}). Then, for $\nu=\nu'$, one readily arrives at the first equation
in (\ref{normgen}).}
\be
\begin{array}{l}
\ds\int\frac{p^2dp}{(2\pi)^3}\left[\vp_{n\nu}^{+}(p)\vp_{m\nu}^{+}(p)-
\vp_{n\nu}^{-}(p)\vp_{m\nu}^{-}(p)\right]=\delta_{nm},\\[3mm]
\ds\int\frac{p^2dp}{(2\pi)^3}\left[\vp_{n\nu}^{+}(p)\vp_{m\nu}^{-}(p)-
\vp_{n\nu}^{-}(p)\vp_{m\nu}^{+}(p)\right]=0.
\end{array}
\label{normgen}
\ee

The representation (\ref{Mmgen}), together with the normalisation
and orthogonality conditions (\ref{normgen}), give the fully
diagonalised Hamiltonian, 
\be 
\hat{\cal H}=\sum_{n,\nu}M_{n\nu}m^\dagger_{n\nu}m_{n\nu}+
O\left(\frac{1}{\sqrt{N_C}}\right), \label{hdgen} \ee provided the
mesonic wave functions obey the bound-state equation, \be \left\{
\begin{array}{l}
[2E_p-M_{n\nu}]\vp_{n\nu}^+(p)=\ds\int\frac{\ds q^2dq}{\ds (2\pi)^3}
[T^{++}_\nu(p,q)\vp_{n\nu}^+(q)+T^{+-}_\nu(p,q)\vp_{n\nu}^-(q)]\\[0cm]
[2E_p+M_{n\nu}]\vp_{n\nu}^-(p)=\ds\int\frac{\ds q^2dq}{\ds (2\pi)^3}
[T^{-+}_\nu(p,q)\vp_{n\nu}^+(q)+T^{--}_\nu(p,q)\vp_{n\nu}^-(q)].
\end{array}
\right.
\label{bsgen}
\ee

On comparing the mass-gap-like equation for the generalised
Bogoliubov transformation (\ref{bsgen}) with the Bethe--Salpeter
equation (\ref{Salpeterlev5}), one can easily see that they
coincide, the only difference being that they are written in
different representations: Eq.~(\ref{Salpeterlev5}) is written in the
quark spins representation, whereas Eq.~(\ref{bsgen}) is written
in the $J^{PC}$ representation. Therefore, it is a matter of
rewriting the Salpeter wave function $\Phi^\pm$ in the proper
representation to go from one equation to the other. Notice that
the Lorentz nature of the confining interaction --- the explicit
form of the matrices $\Gamma$ in Eqs.~(\ref{Sigma01}) and
(\ref{GenericSal}) --- plays no role, defining only the explicit
form the amplitudes $v^{\pm\pm}$.
Similarly, the Hamiltonian approach to the theory can be
developed, following the same lines, for an arbitrary matrix
$\Gamma$.

In the leading order in $N_C$ the Hamiltonian (\ref{hdgen})
describes free stable mesons. The suppressed terms in
(\ref{hdgen}) must involve quark exchange and correspond to the
parts of the Hamiltonian responsible for hadronic decays and
scattering. For example, to consider a decay $A\to B+C$, one
should restore the first sub-leading term in the Hamiltonian,
$\frac{1}{\sqrt{N_C}}\hat{m}_A\;\hat{m}_B^\dagger\;
\hat{m}_C^\dagger$, with the coefficient giving the amplitude of
the corresponding process \cite{2d}. Notice, however, that all
three mesons --- $A$, $B$, and $C$ --- cannot be at rest and one
encounters a problem of boosting mesonic creation/annihilation
operators, which is closely related to the general problem of
Lorentz boosts in such potential models. This important issue lies
beyond the scope of the present paper and deserves special
treatment. Notwithstanding this general problem, we are still left
with an important set of sub-leading physical processes which are
amenable to exact treatment in the context of non-local NJL
Hamiltonians: the set of elastic hadron-hadron scattering at rest
(scattering lengths) where we can define a common vacuum for all
intervening hadrons \cite{pipi}. We leave the issue of vacuum
boosts to future publications.

This completes the matching between the Hamiltonian and the
Bethe--Salpeter approaches to the non-local NJL  models described
in Eq.(\ref{H}). We conclude that {\em diagonalisation of the
Hamiltonian of the theory in the mesonic sector in the leading
order in $N_C$ is equivalent to developing the Bethe--Salpeter
approach to the theory in the ladder/rainbow approximation. In the
Hamiltonian approach, the mass-gap and the bound-state equations
emerge to ensure the cancellation, in the Hamiltonian, of the
anomalous Bogoliubov terms  {\bfseries both} in the quark and in the
mesonic sectors of the theory. The normalisation condition for the
mesonic wave functions $\vp^\pm$ with the minus sign between the
positive- and the negative-energy components follows naturally in
the Hamiltonian approach as the standard normalisation of the
bosonic Bogoliubov amplitudes.}

\subsection{Discussion}

Let us make several concluding remarks concerning the
diagonalisation procedure presented in this section. We started
from the Hamiltonian of a quark model with the instantaneous
interaction parameterised by the quark kernel of an arbitrary, but
necessarily confining, form. As an example for such like models we
have, for instance, QCD in the truncated Coulomb gauge. After
inclusion of self-interaction into the quark fields we obtain
dressed quarks --- the so-called BCS level. Chiral symmetry is
spontaneously broken at this stage, quarks acquire effective mass
and the chiral condensate appears in the vacuum. The next step was
to perform yet another Bogoliubov-type transformation, so as to
build operators which create/annihilate $nJ^{PC}$ mesonic states
in the BCS vacuum. Finally, the rules relating the second
Bogoliubov transformation (which diagonalised, up to
$N_C$-suppressed terms, the quartic part of the Hamiltonian) and
the Bethe--Salpeter equation for bound quark-antiquark states were
deduced. Crucial points of the approach necessary to carry out
this program were:
\begin{itemize}
\item instantaneous inter-quark interaction which allows one to
avoid the problem of the relative inter-quark time and to
formulate a self-consistent Hamiltonian approach to the theory;
\item limit of a large number of colours with the inherent
suppression of all non-planar diagrams (quark exchange);
\item confinement, which allows one to reformulate the theory
entirely in terms of colourless bound states
--- mesons (the crucial conjecture (\ref{anzatz}) will fail for
nonconfining interactions);
\item a nontrivial solution to the mass-gap equation defining the
broken phase of the theory. This phase is characterised by two
concurrencial processes, one defined by the amplitude of creation
of a quark-antiquark pair by means of the $\hat{b}^\dagger
\hat{d}^\dagger$ operator applied to the
vacuum ($\hat{M}^\dagger_{ss'}(\vec{p},\vec{p}')$), and a second
process realised through the \lq\lq borrowing" of $\bar{q} q$
pairs from the chiral condensate via the operator $\hat{d}\hat{b}$
($\hat{M}_{ss'}(\vec{p},\vec{p}')$). Both processes are
fundamentally important for the chiral pion to become massless due
to strong cancellations between the positive- and
negative-energy components of the pionic wave function.
\end{itemize}

The discussed approach is completely insensitive to:
\begin{itemize}
\item the number of spatial dimensions, provided a suitable basis
is built which diagonalises the spatial part of the Hamiltonian.
This is due to the fact that both the mechanism for the separation
of positive from negative-energy components of the mesonic wave
functions and the mechanism for the disentanglement of radial
excitations are universal mechanisms for instantaneous
interactions;
\item the Lorentz nature of confinement, which we considered to
take the simplest form --- $\gamma_0\times\gamma_0$, but which can
be easily generalised to be $\gamma_\mu\times\gamma_\nu$;
\item the form of the potential, which is required only to be
confining and to lead to a finite mass-gap equation. All
power-like potentials (\ref{potential}) with
$0\leqslant\alpha\leqslant 2$ meet these conditions. Among these
we have the linearly rising potential, $\alpha=1$, --- as the most
natural candidate for confinement in QCD, as well as the harmonic
oscillator potential, $\alpha=2$, which is the easiest example for
analytical and numerical studies.
\end{itemize}

Finally, we arrive at the following important conclusions:
\begin{enumerate}

\item the Bethe--Salpeter equation for bound states of quarks and
antiquarks is equivalent to diagonalisation of the quartic term in
the quark Hamiltonian, expressed in terms of the dressed quark
fields. Notice that no new information, at least in the leading
order in $N_C$, is required/appears beyond BCS level, and then
we naturally arrive at the second conclusion that \item the entire
problem is completely defined as soon as the mass-gap equation is
formulated and solved, the chiral angle being the only entity
which one needs to know in order to solve the entire theory. In
particular, \item the second, bosonic, Bogoliubov transformation
is defined via the Salpeter amplitudes $\vp^+$ and $\vp^-$. In the
literature of boson condensation they are usually denoted by $u$
and $v$ amplitudes respectively. Therefore it is not hard to
understand that $\vp^\pm$ play simultaneously the role of the
mesonic wave functions and satisfy the normalisation condition
which is proper of the usual Bogoliubov self-consistency condition
$u^2-v^2=1$. Thence these two amplitudes can be parameterised as
$u=\cosh\chi$ and $v=\sinh\chi$, with $\chi$ being the mesonic
Bogoliubov angle. Therefore, for the generalised mesonic
transformation, $\cosh\chi$ and $\sinh\chi$ are given by the sets
of the positive- and negative-energy components $\{\vp^+_{n\nu}\}$ and
$\{\vp^-_{n\nu}\}$ which are completely defined by the chiral angle
(for example, for the chiral pion we have
$\vp_\pi^+=-\vp_\pi^-={\cal N}_\pi\sin\vp_p$).
\end{enumerate}

{\em In short, solving the mesonic Bethe-Salpeter equation is
tantamount to finding the bosonic Bogoliubov transformation to
fully diagonalise --- in the leading order in $N_c$ --- any non-local
NJL Hamiltonian}.

This ends the discussion of the diagonalisation of non-local NJL
models. In the next section, we generalise the results of this
section to the case of multiple vacuum states --- replicas.

\section{Vacuum replicas in potential quark models}

\subsection{Multiple solutions to the mass-gap equation}

As noticed above, the mass-gap equation is equivalent in momentum
space to a Schr{\" o}dinger-like equation, with the dressed quark
dispersive law playing the role of the effective potential (see
the example of Eq.~(\ref{mgho})). This suggests that the solution
to the mass-gap equation might not be unique. Such multiple
solutions were discovered for the harmonic oscillator potential in
\cite{Orsay,Lisbon}. Recently a detailed analysis was performed
for the general form of the power-like confining potential,
Eq.~(\ref{potential}), and the existence of an infinite tower of
solutions to the corresponding mass-gap equation was proved
\cite{replica4}. Let us enumerate the main statements one can make
concerning the mass-gap equation:
\begin{itemize}
\item in order to provide a nontrivial solution to the mass-gap
equation the dressed quark dispersive law must become negative in
the infrared region $p\to 0$. Since the chiral symmetry is broken
only for such nontrivial chiral angles, then this property of
$E_p$ is absolutely necessary for chiral symmetry breaking;
\item once a nontrivial solution $\vp_p$ to the mass-gap equation
is found, it defines the wave function of the pion --- the
Goldstone boson --- as $\psi={\cal N}_\pi\sin\vp_p$;
\item the slope of the chiral angle at the origin defines the
scale of the chiral symmetry breaking for the given solution;
steeper solutions correspond to less broken symmetry and to a
sharper behaviour of $E_p$ at the origin; \item the mass-gap
equation for any single-parameter confining potential leads to an
infinite number of solutions for the chiral angle.
\end{itemize}

To exemplify the appearance of the replicas, let us draw the
following qualitative picture. For the trivial chiral angle
$\vp_p=0$ there is no dressing of quarks, the quark dispersive law
being just $E_p=p$ for all momenta. With such $E_p$, the effective
potential in the Schr{\" o}dinger equation (\ref{mgho}) (and
similarly for other forms of the confining potential), $V_{\rm
eff}(p)=2E_p(p)$, does not possess a single bound state with a
zero eigenvalue --- therefore, no nontrivial solutions to the
mass-gap equation do exist. We encounter here the well-known
problem of the constituent quark models. Indeed, as it was
discussed above, the mass-gap equation plays the role of the
bound-state equation for the pion at rest. Upon substituting the
free quark dispersive law, $E_p=p$, into the mass-gap
equation --- see for example, Eq.~(\ref{mgho}) --- we are led to the
usual Schr{\" o}dinger equation for the pion,
$[2p+V(r)]\psi_\pi=M_\pi\psi_\pi$, which is characteristic of
potential constituent quark models. The pion mass coming out of
this equation appears to be of order of the confining interaction
scale, $M_\pi\sim 400\div 500MeV$, which is several times larger
than the experimental value of $140MeV$ and such $M_\pi$ {\em does
not} vanish in the chiral limit. In the bound-state equation, this
problem is known to be solved by the presence of the
negative-energy component $\vp_\pi^-$ of the pionic wave function
which happens to be of the same order of magnitude as the
positive-energy component $\vp_\pi^+$. Strong cancellations
between the two components of the pion wave function bring the
pion mass to zero. At BCS level, the solution of this problem
comes from the actual form of the dressed quarks dispersive law
$E_p$ which becomes negative at small momenta, and cancels the
positive contribution of the confining potential allowing for the
pion mass to vanish. In other words: the peculiar behaviour of the
dressed quark dispersive law is {\em both a necessity and a direct
consequence of the chiral symmetry breaking}. Suppose now that we
can parameterise this small-$p$ negative contribution to $E_p$ by
a mass scale $\mu$. For the power-like potentials
(\ref{potential}), it is clear that such a contribution has to be
proportional to $K_0^{1+\alpha}$ (see the definition of $E_p$
through the auxiliary functions $A_p$ and $B_p$, Eq.~(\ref{Ep}),
as well as Eqs.~(\ref{A}) and (\ref{B})). For dimensional reasons,
$E_p(p=0)=-{\rm const \frac{K_0^{1+\alpha}}{\mu^\alpha}}$ (for
example, for $\alpha=2$, from Eq.~(\ref{Epharm}) with $m=0$ and
$\vp_p\mathop{\approx}\limits_{p\to 0}\frac{\pi}{2}-\frac{p}{\mu}$
one readily finds that $E_p(p=0)=-\frac32\frac{K_0^3}{\mu^2}$).
Thus, for a sufficiently small $\mu =\mu_0$ the effective
potential $V_{\rm eff}(p|\mu)$ becomes binding enough to produce
an eigenstate with a zero eigenvalue. Then $\mu_0$ defines the scale
of the chiral symmetry breaking in the new vacuum and the
corresponding chiral angle behaves as
$\vp_0(p)\mathop{\approx}\limits_{p\to
0}\frac{\pi}{2}-\frac{p}{\mu_0}+\ldots$. This is the ground-state
solution which corresponds to the BCS vacuum of the theory.
Decreasing the scale $\mu$, one can reach a situation where a
second zero eigenvalue bound state appears for the potential
$V_{\rm eff}(p|\mu)$. According to general quantum mechanical
theorems, this solution must contain one knot. This is the first
replica vacuum. Continuing this procedure, we can build the whole
infinite tower of replicas for a given potential. In other words,
for any given value of the scale $\mu$ one has a set of orthogonal
eigenstates (for the potential $V_{\rm eff}(p|\mu)$) with positive
and negative eigenvalues. To search for the $n$th replica, it is
sufficient to vary $\mu$, thereby shifting the whole tower of
eigenstates up or down until the appearance of the sought $n$th
zero eigenvalue. Therefore, different replicas are eigenstates in
different potentials.

The easiest way to prove such a picture is to evaluate the
quasi-classical Bohr-Sommerfeld integral for such an eigenvalue
problem. Since we consider a one-scale confining interaction,
then, from dimensional reasons, it is clear that the kinetic and
the potential energies, in momentum space, behave as
$K^{1+\alpha}r^\alpha$ and $K_0^{1+\alpha}/p^\alpha$,
respectively.  Consequently the corresponding WKB integral depends
logarithmically on the scale $\mu$, which plays the role of the
cut-off, $I_{\rm WKB}\propto \ln\frac{K_0}{\mu}$ \cite{replica4}.
Therefore, the quasi-classical quantisation condition,
$I_{WKB}=2\pi n$, can be fulfilled for any $n$, provided the
corresponding scale $\mu_n$ is small enough. For example, for the
harmonic oscillator potential in $D$ dimensions, the approximate
dependence of the scale $\mu$ on the index of the replica
$n$ can be easily found using the aforementioned WKB method to be,
\be
\mu_n={\rm const}\times K_0\exp{\left(-\frac{2\pi n}{\sqrt{D(D-2)}}\right)},
\ee
so that, in any dimension $D>2$, the number of replicas is,
indeed, infinite. The boundary case of $D=2$ contains a
singularity, for $n\neq 0$, so there are no replica solutions in
the harmonic oscillator potential in two dimensions, as was found
in \cite{replica1} and discussed in detail in \cite{replica5}.

Of course, the procedure of building replicas drawn above should
be understood as a simplified qualitative method, since the scale
$\mu$ is not a free parameter but it has to be defined
self-consistently with the form of the chiral angle. Notice also
that, for confining potentials defined through more than one
parameter, the proof given above does not hold, since the WKB
integral may be regularised by the second scale, instead of the
parameter $\mu$. Then the number of replicas becomes finite or
they do not exist at all. Notice, however, that, in the general
case, it is much harder not to have replicas than to have them,
and the expectation of existence of more than one solution for
nonlinear mass-gap equations is not only physically attractive but
also quite natural from the mathematical point of view.

\subsection{The problem of the tachyon}

In this subsection, we address the problem already touched upon in
the paper \cite{replica1} --- namely, the problem of the tachyon
which appears in the excited vacua. Indeed, it was numerically
found that, in the chiral limit, the chiral condensate changes the
sign from replica to replica remaining negative for even states
(the ground-state BCS vacuum being the lowest representative of
the even states) and becomes positive for all odd states, starting
with the first replica \cite{replica1,replica4}. The universal
status of this rule becomes clear from the asymptotic behaviour of
the chiral angle (\ref{asym}) and the fact that for each next
replica the chiral angle possesses an extra knot and, therefore,
it approaches, when $p\rightarrow \infty$, the $\vp_p=0$ asymptote
from the opposite half-plane, as compared to the previous replica.
Then, switching on a small quark mass, we arrive at the
Gell-Mann--Oakes--Renner relation (\ref{GMOR}), in which the
chiral condensate on the r.h.s. can be calculated in the limit
$m=0$ \ftnote{5}{Beyond the chiral limit the quark condensate
acquires an infinite contribution coming from the trace of the
free-particle Green's function which should be subtracted. The
term of the zeroth order in $m$ in the properly regularised
condensate obviously coincides with the value calculated in the
chiral limit.}, and, for positive values of
$\langle\bar{q}q\rangle$, the pion becomes a tachyon, $M_\pi^2<0$.
As always, the presence of a tachyon means that the lowest state
--- the vacuum
--- is chosen improperly, and there should be a more preferable vacuum.
It is easy to see that, for odd replicas, this is indeed the case.

To demonstrate this, let us consider the exact chiral limit and
the most general Valatin-Bogoliubov transformation from the
trivial vacuum $|0\rangle_0$ to the generalised $\theta$-vacuum
$|\theta\rangle$ which can be written as,
\be
|\theta\rangle=S_\theta|0\rangle_0,\;\;
S_\theta=\exp{[Q_\theta^\dagger-Q_\theta]},\;\; Q_\theta^\dagger
=\frac12\sum_{\alpha}\sum_{ss'}\int \frac{d^3p}{(2\pi)^3}\vp_p
\mathfrak{M}_{ss'}[\theta]\hat{b}_{\alpha s}^\dagger(\vec{p})\hat{d}_{\alpha
s'}^\dagger(-\vec{p}),
\label{gBV}
\ee
where $\vp_p$ is the chiral angle and
\be
\mathfrak{M}[\theta]=\mathfrak{M}_{^3P_0}\cos\theta
+i\mathfrak{M}_{^1S_0}\sin\theta.
\ee

The $^3P_0$ matrix
$\mathfrak{M}_{^3P_0}=(\vec{\sigma}\hat{\vec{p}})i\sigma_2$ is
studied in detail in \cite{Lisbon} and the $^1S_0$ matrix
$\mathfrak{M}_{^1S_0}\equiv \mathfrak{M}_\pi=-i\sigma_2$. For
$\theta=0$ Eq.~(\ref{gBV}) reproduces the standard definition of
the BCS vacuum $|0\rangle$ \cite{Lisbon}. When
$\theta=\frac{\pi}{2}$ we obtain:
\be
\left|\frac{\pi}{2}\right\rangle=S_{\pi/2}|0\rangle_0=
\exp{\left[i\int
d^3x\bar\psi(x)\gamma_0\gamma_5\psi(x)\right]}|0\rangle_0=
\exp{[i\hat{Q}_5]}|0\rangle_0,
\label{q5q}
\ee
with the operator $\hat{Q}_5$ being responsible for translations
along the $\theta$ direction. The angles $\theta$ and $\vp_p$ are independent
quantities. For a given solution of the mass-gap equation,
$\vp_p$, when $\theta=\pi$ we get $S_\pi[\vp_p]=S_0[-\vp_p]$,
which, as discussed above, is the usual pseudo-unitary $^3P_0$
operator for the chiral condensation. Thus, translation from $\theta=0$
to $\theta=\pi$ along the $\theta$ direction is tantamount to a
chiral angle transformation of $\vp_p\to-\vp_p$. We could have
started with $-\vp_p$ --- also a solution to the mass-gap
solution for $m_q=0$ --- to arrive, in the end of the
$\theta$-journey, at $\vp_p$. Notice that, in general, the state
$|\theta\rangle$ does not have any definite parity ---
we have to consider mixtures of different $\theta$-vacua in order to
build a state with a given definite parity. For instance,
$|\theta\rangle+|-\theta\rangle$ has the parity plus. In the
special case of $\theta=\pi$, $S_\pi=S_{-\pi}$ and the state
$|\theta =\pi\rangle$, as well as $|\theta =0\rangle$, has positive parity.

Let us start with a given $\vp_p$ and put, for a moment,
$\theta=0$. If we now plot the vacuum energy as a function of the
chiral condensate $\Sigma=\langle\bar{q}q\rangle$, then $\Sigma=0$
corresponds to the local maximum, whereas the stable minimum is
provided by a nonzero value of $\Sigma$, easily related to the
corresponding chiral angle,
\be
\Sigma=-\frac{N_C}{\pi^2}\int^{\infty}_0 dp\;p^2\sin\vp_p.
\label{Sigma1}
\ee
Next, we can move along the $\theta$-valley. Since, in the chiral
limit, the charge $\hat{Q}_5$ commutes with the Hamiltonian of the
theory, $[\hat{Q}_5\hat{H}]=0$, then the vacuum energy is
degenerate for all $\theta$'s, so that we are free to choose any
value for $\theta$. Such a form of the vacuum energy as a function
of two variables, $\Sigma$ and $\theta$, is known as the Mexican
hat. Incidentally, we can find, starting from expression
(\ref{q5q}), the chiral pion Salpeter amplitude
$\vp_\pi^+(p)=-\vp_\pi^-(p)$, which is just the c-number multiplying
the anomalous part of $\hat{Q}_5$ \cite{emilioResina}. It is then
clear that the pion Salpeter amplitude is given by $\sin\vp_p$,
independently of any consideration for a given quark kernel,
provided it supports the mechanism of spontaneous chiral symmetry
breaking.

\begin{figure}[t]
\centerline{\epsfig{file=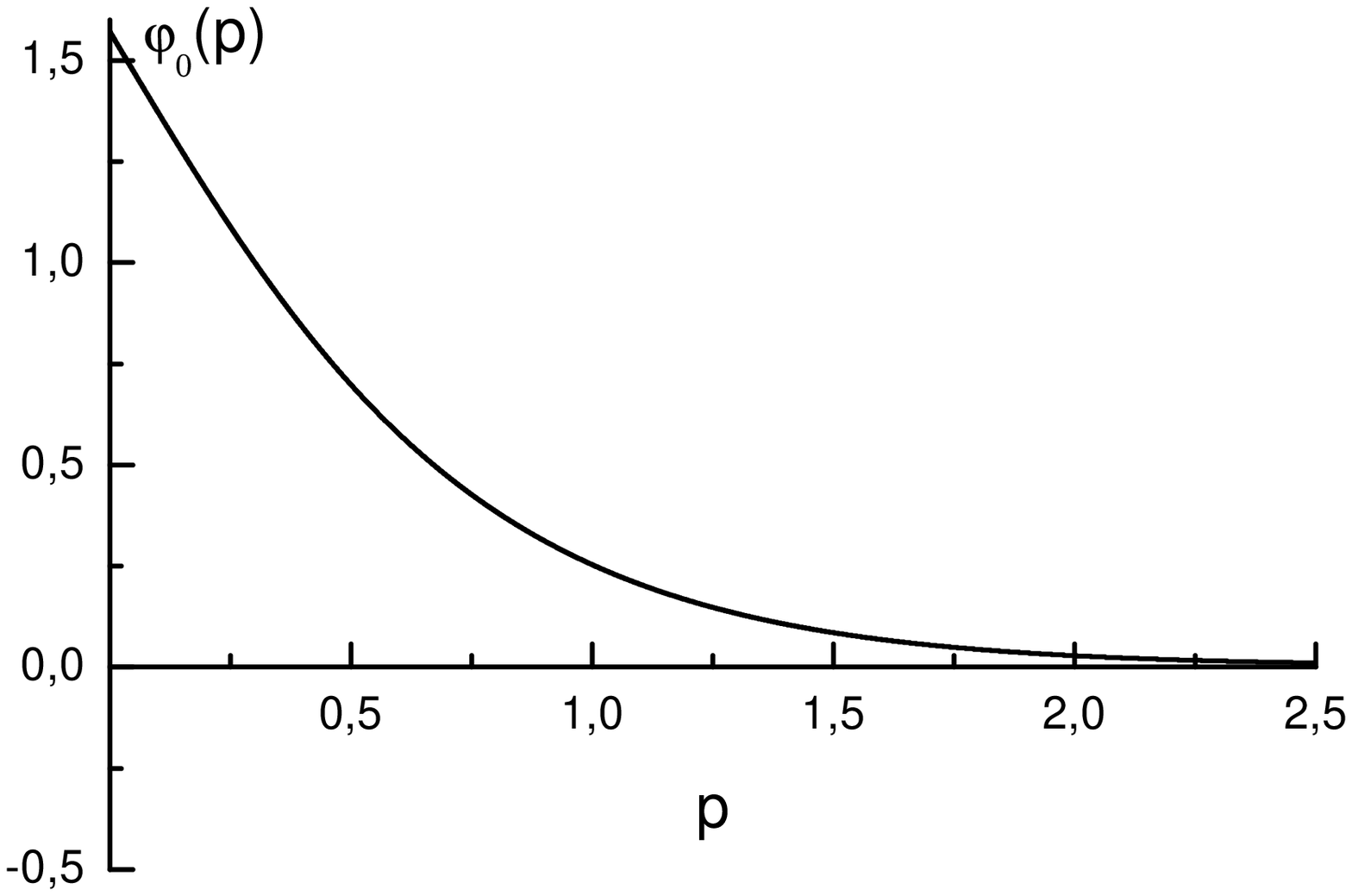,width=8cm}
\epsfig{file=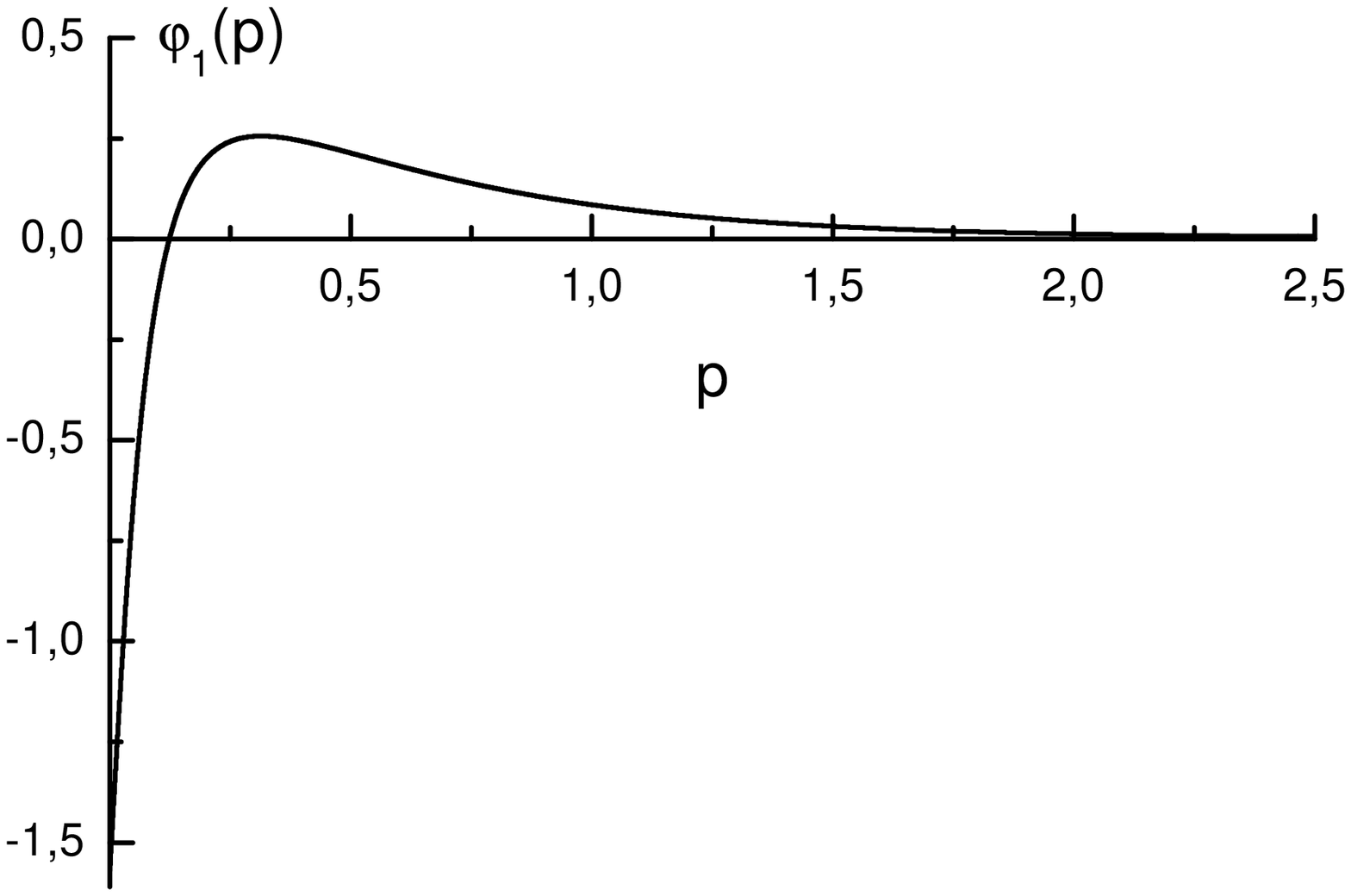,width=8cm}}
\caption{The ground-state (the left plot) and the first replica
(the right plot) solutions of the mass-gap equation (\ref{diffmge})
after the redefinition of the chiral angle sign for odd replicas.
The current quark mass is $m=0.01K_0$.
The momentum $p$ is given in the units of $K_0$.}
\end{figure}

If now a small quark mass is introduced, then the vacuum energy
acquires a contribution of the chirally non-invariant term
proportional to the quark mass $m$ and, as a result, the whole
picture gets tilted
--- the state with the negative sign of the chiral condensate
being energetically preferable as compared to the state with the
positive condensate. These two states differ by their $\theta$
coordinate --- one of them corresponds to $\theta=0$, whereas the
other one has $\theta=\pi$. The choice of the chiral angle adopted
before, with $\vp_p(0)=\frac{\pi}{2}$, selects the vacuum
$\theta=0$ as the true vacuum of the theory $|0\rangle$. This is
indeed the case for the ground BCS vacuum, as well as for all even
replicas, with $\Sigma<0$. The pion, which is not tachyon in these
vacua, is responsible for the vacuum energy increase during the
rotation in the angle $\theta$ from $|0\rangle$ to $|\pi\rangle$.
For odd replicas, the chiral condensate changes the sign, so that
the state with $\theta=\pi$ must acquire a lower energy than the
one with $\theta=0$. As a result, in odd replicas, the pion
becomes the tachyon responsible for the energy decrease during the
same rotation form $|0\rangle$ to $|\pi\rangle$, the latter
representing now the true vacuum. Therefore, for odd replicas, we set
$\vp_p(0)= -\frac{\pi}{2}$ instead of $\frac{\pi}{2}$. This
choice will ensure that the chiral angle, when the momentum $p$ goes to
infinity, will approach zero from above. Then the chiral
condensate (\ref{Sigma1}) changes the sign, and so does the pion
mass squared
--- the pion mass becomes real. As discussed before, the
definition of the chiral angle admits such a change and, as a
result, we arrive at two different classes of solutions to the
mass-gap equation: even solutions, which start from
$\frac{\pi}{2}$ at $p=0$ and approach the free limit at large
$p$'s from above, and odd solutions which also approach their
large-$p$ asymptote from above but, at $p=0$, start from
$-\frac{\pi}{2}$. This solves the problem of the tachyon for odd
replicas, making the latter normal vacuum states with the
possibility of building the spectrum of hadrons above them.

As an example,  we give in Fig.~2, the profiles of the
ground-state and the first replica solutions to the mass-gap
equation (\ref{diffmge}) for the oscillator-type potential (the
case of the generalised power-like confining potential
(\ref{potential}) is studied in detail in \cite{replica4}). As
discussed before, the sign of the chiral angle for the first
replica vacuum is reversed.

In Fig.~3, we represent the solutions of the bound-state problem
for the pion with the harmonic oscillator confining potential,
Eq.~(\ref{hop}), for the two lowest solutions to the mass-gap
equation (\ref{diffmge}) depicted in Fig.~2
--- for the ground BCS vacuum (the left plot in Fig.~3) and for the first replica
with the reversed sign of the chiral angle (the right plot in
Fig.~3). It is clearly seen from Fig.~3 that the pionic wave
functions are indeed very close to $\pm\sin\vp_p$ (see Fig.~2)
with the corrections given by the formula (\ref{vppm}). Notice
that, after the change $\vp_p\to-\vp_p$ performed for odd
replicas, the \lq\lq$+$" and the \lq\lq$-$" components of the
mesonic wave function substitute each other, so that one needs
either to redefine the whole set of the operator transformations
used to diagonalise the Hamiltonian of the theory in the mesonic
sector or, which is more economical, simply to rename $\vp^+$ to
$\vp^-$ and {\em vice versa}.

\begin{figure}[t]
\centerline{\epsfig{file=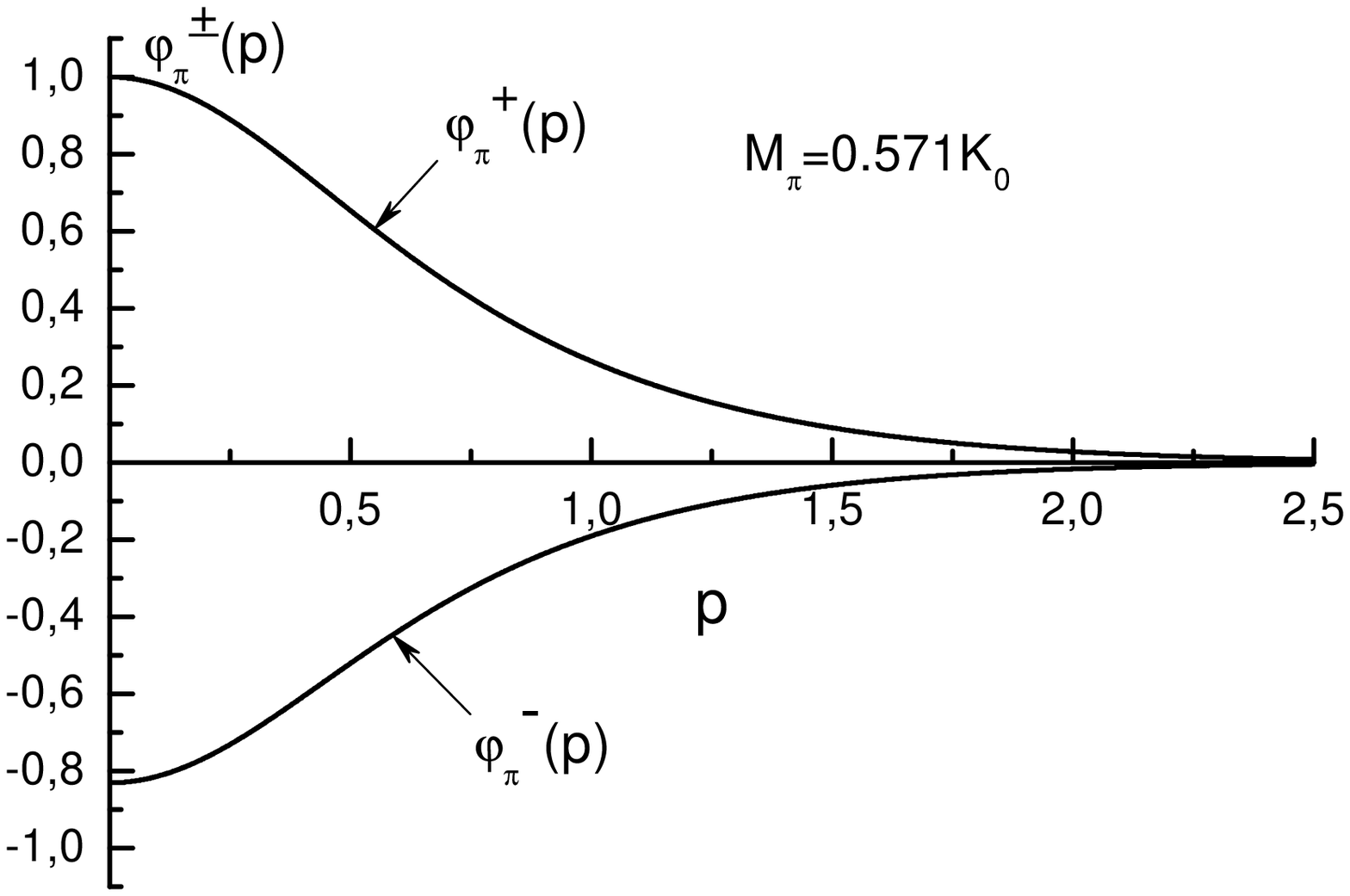,width=7.5cm}\hspace*{10mm}\epsfig{file=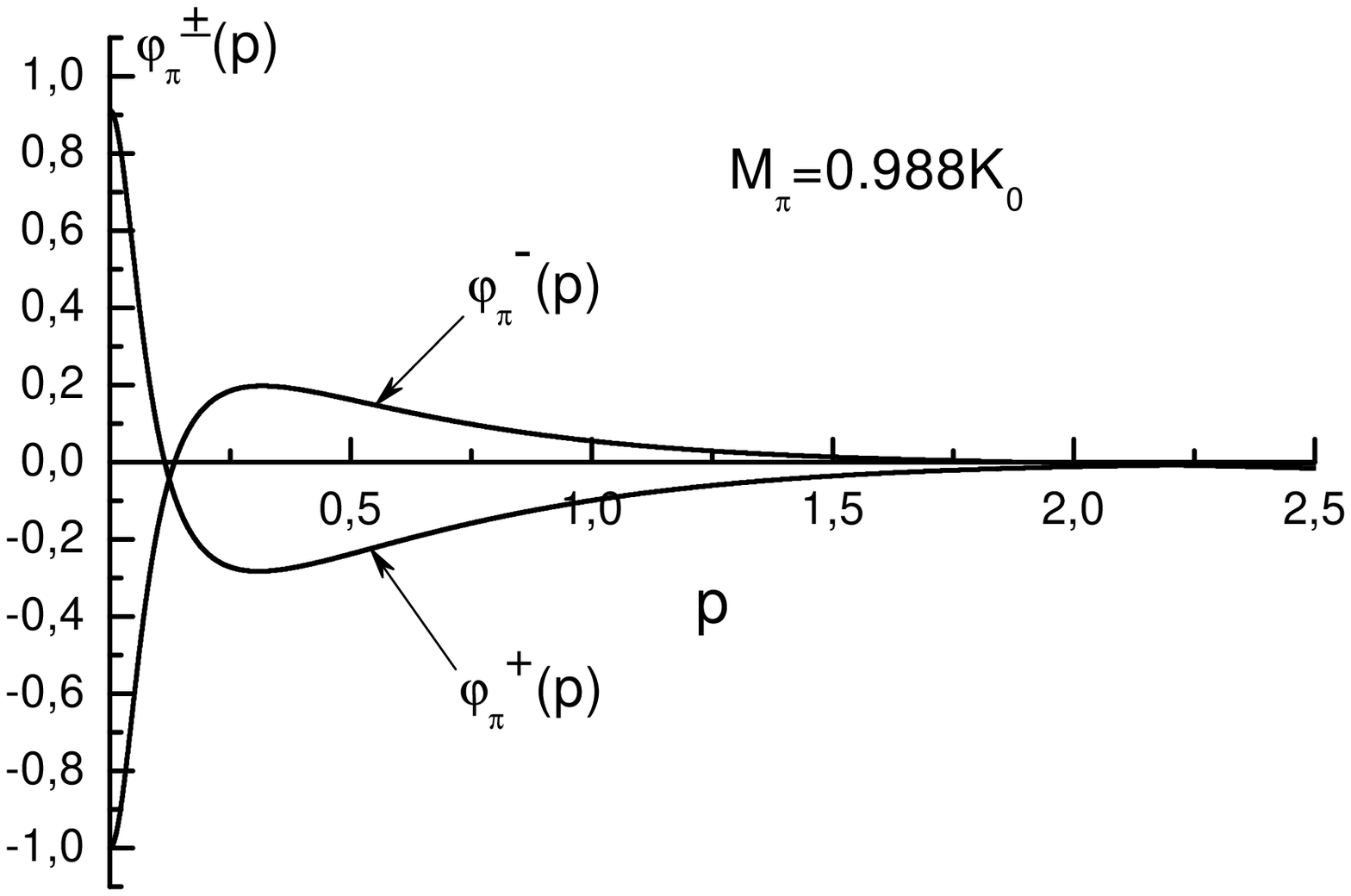,width=7.5cm}}
\caption{Solutions of the bound-state problem for the pion, Eq.~(\ref{hop}), in the BCS vacuum
(the left plot) and in the first replica with the reversed sign of the chiral angle
(the right plot). The overall normalisation factor is omitted, the
largest component being fixed equal to unity at $p=0$.
The current quark mass is $m=0.01K_0$. The momentum $p$
is given in the units of $K_0$.}
\end{figure}

\subsection{Hadronic spectrum in the replica vacua}

It is concluded, in the second section, that any NJL-type
Hamiltonian model with an arbitrary confining quark kernel is
completely defined at BCS level, as soon as the chiral angle
is built. If the large-$N_C$ limit is assumed, then all
approximations become controllable and the Hamiltonian of the
theory admits complete diagonalisation in terms of the compound
mesonic operators. We found that the appropriate basis for such a
diagonalisation was provided by the set $\{n,\;J^{PC}\}$ and each
mesonic state was described by a pair of wave functions $\vp^\pm$.
In this section, following the papers
\cite{replica1,replica2,replica4}, we gave evidence of existence
of excited solutions to the mass-gap equation --- the vacuum
replicas. Now we address the question concerning the spectrum of
hadrons in the replica vacua.

As argued in the previous subsection, the pion is massless in the
chiral limit for any replica vacuum and, with the properly defined
chiral angle, it acquires a small real mass beyond the chiral
limit. As far as highly excited mesonic states are concerned, the
chiral symmetry is restored in this part of the spectrum and the
hadrons' properties become insensitive to the details of the
vacuum in which they are created since, for high excitations, the
negative-energy components $\vp^-_{n\nu}(p)$ of the wave functions
are negligible and the Bethe--Salpeter bound-state equation
(\ref{bsgen}) amounts to a quantum-mechanical Schr{\" o}dinger equation,
\be
[2E_p+V(r)]\vp^+_{n\nu}(p)=M_{n\nu}\vp^+_{n\nu}(p),\quad
E_p=\sqrt{p^2+m^2}.
\label{Sal}
\ee
Consequently, no cancellations
between the positive- and negative-energy components take place
anymore, as it happens to the chiral pion and leads to the small
pion mass of $140MeV$, as opposed the value of about $400\div
500MeV$ which would follow from Eq.~(\ref{Sal}). As a result, in
the vacua with \lq\lq less" broken chiral symmetry (higher
replicas), the masses of mesons are pushed up, with the exception
of the chiral pion, whose mass, in the chiral limit, is kept equal
to zero by the requirement of the chiral symmetry. Beyond the
chiral limit, the pion mass also increases with the index
of the replica (see Fig.~3 and compare the pion mass of $0.57K_0$,
in the BCS vacuum, with the value of $0.988K_0$, for the first
replica).

Therefore, we conclude that each mesonic state can be
characterised by an extra \lq\lq quantum number" --- the index
of the vacuum replica in which it was created. Then the
full diagonalised Hamiltonian of the theory takes the form: 
\be
\hat{\cal H}=\sum_{n,\nu,{\cal N}}M_{n\nu{\cal
N}}m^\dagger_{n\nu{\cal N}}m_{n\nu{\cal
N}}+O\left(\frac{1}{\sqrt{N_C}}\right), 
\label{hdgen2} 
\ee 
${\cal N}$ being the replica label, which is the generalisation
of Eq.~(\ref{hdgen}) for the multi-vacuum case.

\section{Conclusions}

In this paper, we have shown how the generalised NJL-type
Hamiltonian model with a nonlocal confining quark kernel can be
fully diagonalised in the sector of physically observable hadronic
states. We extend this result to include the multiple vacuum
states recently found to exist in models of such a type and which
are very likely to exist in real QCD. The main result of the paper
is the fully diagonalised Hamiltonian (\ref{hdgen2}). Similarly to
the case of a path integral with a multi-minimum function in the
exponent, the Hamiltonian (\ref{hdgen2}) sums the contributions of
all minima, with the corresponding weight --- the masses of mesons
in the given vacuum --- being minimal for the lowest vacuum state
and increasing for higher vacua, thus suppressing the contribution
of the corresponding replicas to the full partition function of
the theory. We study the form of the chiral angle profiles
defining the excited vacua and find that for an even (odd) replica
to decay to a lower even (odd) one it is sufficient to change the
chiral angle in the finite region in momentum $p$, since both
asymptotes for both replicas of the same parity coincide:
$\vp_p(p=0)=\frac{\pi}{2}\;\left(-\frac{\pi}{2}\right)$,
$\vp_p(p\to\infty)=0$. On the contrary, for the decay of an odd
replica to an even one, or {\em vice versa}, the chiral angle has
to be changed dramatically in the infrared region, since the
low-momentum behaviour of even and odd replicas is quite different.

In the paper \cite{replica2} a quantum field theory perturbative method was
developed. It used the replica-quark-antiquark vertex
parameterised by a small difference between the chiral angles of
the ground-state vacuum and the replica to evaluate the quark self-energy
shift due to the replica presence. This method can be easily
extended to the case of an infinite number of replicas, but notice
that the sets of even and odd states have to be considered
separately. As discussed above, only transformations of the chiral
angle between states with the same parity are small and can be
considered perturbatively.
We conclude, therefore, that two sets of perturbative
approaches to replicas should be built --- for even and for odd replicas,
independently.
On the contrary, a transition between
states with opposite parities, for example, the decay of the first
replica to the ground BCS vacuum, involves the global
transformation of the chiral angle generated by the pseudoscalar
pionic operator $\hat{Q}_5$ and results in a burst of pions with
a huge energy release. The same transition in the opposite
direction --- excitation of the replica
--- requires an external global source which is not
inherent to the model. Building of such a source is an important
problem in the theory of replicas and it will be the subject of
future publications.

\begin{acknowledgments}
One of the authors, A. Nefediev, is grateful to P. Bicudo and Yu.
S. Kalashnikova for many fruitful discussions and would like to
thank the staff of the Centro de F\'\i sica das Interac\c c\~oes
Fundamentais (CFIF-IST) for cordial hospitality during his stay in
Lisbon, where this work was originated, and to acknowledge the
financial support of the grant NS-1774.2003.2, as well as of the
Federal Programme of the Russian Ministry of Industry, Science and
Technology No 40.052.1.1.1112.
\end{acknowledgments}

\appendix
\section{The pion Bethe-Salpeter equation}

In this appendix we derive the Bethe-Salpeter equation for the pion
for the generic form of the potential and
for the Lorentz structure of the confinement being $\gamma_0\times\gamma_0$.
We generalise the method suggested in the paper \cite{BG} for two-dimensional QCD.
We start from Eq.~(\ref{GenericSal}) for the mesonic Salpeter amplitude and
define a matrix mesonic wave function,
\be
\Psi(\vec{p};M_\pi)=\int\frac{dp_0}{2\pi}S(\vec{p},p_0+M_\pi/2)\chi(\vec{p};M_\pi)S(\vec{p},p_0-M_\pi/2).
\ee

We also present the Dirac projectors (\ref{Feynman}) in the form:
\be
\Lambda^\pm(\vec{p})=T_pP_\pm T_p^\dagger,\quad
P_\pm=\frac{1\pm\gamma_0}{2},\quad
T_p=\exp{\left[-\frac12(\vec{\gamma}\hat{\vec{p}})\left(\frac{\pi}{2}-\vp_p\right)\right]},
\ee
and introduce a modified wave function
$\tilde{\Psi}(\vec{p};M_\pi)=T^\dagger_p\Psi(\vec{p};M_\pi)T^\dagger_p$.
Equation for this new matrix wave function following from
Eq.~(\ref{GenericSal}), with $\Gamma=\gamma_0$, reads:
\be
\tilde{\Psi}(\vec{p};M_\pi)=-\int\frac{d^3q}{(2\pi)^3}V(\vec{p}-\vec{q})\left[
P_+\frac{T_p^\dagger
T_q\tilde{\Psi}(\vec{q};M_\pi)T_qT_p^\dagger}{2E_p-M_\pi}P_-
+P_-\frac{T_p^\dagger
T_q\tilde{\Psi}(\vec{q};M_\pi)T_qT_p^\dagger}{2E_p+M_\pi}P_+
\right].
\label{PHI}
\ee
It is clear that a solution of Eq.~(\ref{PHI}) has the form,
\be
\tilde{\Psi}(\vec{p};M_\pi)=P_+AP_-+P_-BP_+,
\label{AB}
\ee
and, due to the obvious orthogonality property of the projectors
$P_\pm$, $P_+P_-=P_-P_+=0$, only matrices anti-commuting with the
matrix $\gamma_0$ contribute to $A$ and $B$. The set of such
matrices is
$\{\gamma_5,\gamma_0\gamma_5,\vec{\gamma},\gamma_0\vec{\gamma}\}$
which can be reduced even more, up to $\{\gamma_5,\vec{\gamma}\}$,
since the matrix $\gamma_0$ can be always absorbed into projectors
$P_\pm$. For the case of the chiral pion only $\gamma_5$
contributes and one has:
\be
A_\pi=\gamma_5\vp^+_\pi(p),\quad
B_\pi=-\gamma_5\vp_\pi^-(p),
\label{AB2}
\ee
where the signs and the coefficients are chosen such that to comply with the
definitions (\ref{Mmpi}) and (\ref{vppi}). It is an easy task now
to extract the amplitudes $T_\pi^{\pm\pm}$ (see Eq.~(\ref{bsp}))
from Eq.~(\ref{PHI}) using the explicit form of $\tilde{\Psi}_\pi(\vec{p};M_\pi)$
and the operator $T_p$. They read:
\be
\begin{array}{lcr}
T_\pi^{++}(p,q)=T_\pi^{-+}(p,q)&=&-\ds\int d\Omega_q V(\vec{p}-\vec{q})
\left[\cos^2\frac{\vp_p-\vp_q}{2}-\frac{1-(\hat{\vec{p}}\hat{\vec{q}})}{2}\cos\vp_p\cos\vp_q\right],\\[5mm]
T_\pi^{+-}(p,q)=T_\pi^{--}(p,q)&=&\ds\int d\Omega_q V(\vec{p}-\vec{q})
\left[\sin^2\frac{\vp_p-\vp_q}{2}+\frac{1-(\hat{\vec{p}}\hat{\vec{q}})}{2}\cos\vp_p\cos\vp_q\right].
\end{array}
\label{pia}
\ee

For the harmonic oscillator potential, and with the amplitudes (\ref{pia}),
Eq.~(\ref{bsp}) reproduces Eq.~(\ref{hop}).

In the chiral limit, $\vp_\pi^+(p)=-\vp_\pi^-(p)=\vp_\pi(p)$, so that the
bound-state equation (\ref{bsp}) reduces to a single equation,
\be
2E_p\vp_\pi(p)=\int\frac{q^2dq}{(2\pi)^3}[T_\pi^{++}(p,q)-T_\pi^{+-}(p,q)]\vp_\pi(q)=
-\int\frac{d^3q}{(2\pi)^3}V(\vec{p}-\vec{q})\vp_\pi(q),
\ee
or, in the coordinate space, one arrives at the Schr{\" o}dinger-like equation,
\be
[2E_p+V(r)]\vp_\pi=0.
\ee

It is also instructive to derive the bound-state equation for the pionic matrix wave function
$\Psi(\vec{p};M_\pi)$. It follows directly from the representation (\ref{PHI}) after a
simple algebra (see \cite{2d} for the detailed discussion of the matrix wave function formalism
in two-dimensional QCD),
$$
M_\pi\Psi(\vec{p};M_\pi)=[(\vec{\alpha}\vec{p})+\gamma_0m]\Psi(\vec{p};M_\pi)+
\Psi(\vec{p};M_\pi)[(\vec{\alpha}\vec{p})-\gamma_0m]\hspace*{5cm}
$$
\be
+\int\frac{d^3q}{(2\pi)^3}V(\vec{p}-\vec{q})\left\{\Lambda^+(\vec{q})\Psi(\vec{p};M_\pi)\Lambda^-(-\vec{q})-
\Lambda^+(\vec{p})\Psi(\vec{q};M_\pi)\Lambda^-(-\vec{p})\right.
\label{matrix}
\ee
$$
\hspace*{5cm}\left.-\Lambda^-(\vec{q})\Psi(\vec{p};M_\pi)\Lambda^+(-\vec{q})+
\Lambda^-(\vec{p})\Psi(\vec{q};M_\pi)\Lambda^+(-\vec{p})\right\}.
$$

The explicit form of $\Psi(\vec{p};M_\pi)$ through the components $\vp_\pi^\pm$ and the
chiral angle can be found easily from Eqs.~(\ref{AB}), (\ref{AB2}),
\be
\Psi(\vec{p};M_\pi)=T_p\left[P_+\gamma_5\vp_\pi^+-P_-\gamma_5\vp_\pi^-\right]T_p=
\gamma_5G_\pi+\gamma_0\gamma_5T_p^2F_\pi,
\label{exp}
\ee
where $G_\pi=\frac12(\vp_\pi^+-\vp_\pi^-)$, $F_\pi=\frac12(\vp_\pi^++\vp_\pi^-)$.

Let us multiply the matrix bound-state equation (\ref{matrix}) by $\gamma_0\gamma_5$, integrate its
both parts over $d^3p$, and, finally, take the trace. The resulting equation reads:
\be
M_\pi\int\frac{d^3p}{(2\pi)^3}F_\pi\sin\vp_p=2m\int\frac{d^3p}{(2\pi)^3}G_\pi.
\label{GMOR2}
\ee

It is easy to recognise the Gell-Mann--Oakes--Renner relation (\ref{GMOR}) in
Eq.~(\ref{GMOR2}). Indeed,
using the explicit form of the pionic wave function beyond the chiral limit,
Eq.~(\ref{vppm}), one can
see that $G_\pi=(\tilde{\cal N}_\pi/\sqrt{M_\pi})\sin\vp_p$ and
$F_\pi=(\tilde{\cal N}_\pi\sqrt{M_\pi})\Delta_p$ which, after substitution to
Eq.~(\ref{GMOR2}), give the sought relation,
\be
M_\pi^2\left[\frac{N_C}{\pi^2}\int_0^\infty dp\; p^2\Delta_p\sin\vp_p\right]=-2m\langle\bar{q}q\rangle,
\label{GMOR3}
\ee
where the definition of the chiral condensate (\ref{Sigma1}) was used.
\end{document}